\documentclass[preprintnumbers,prd,twocolumn,showpacs,floatfix,
preprintnumbers,letterpaper,nofootinbib,superscriptaddress]{revtex4-1}
\usepackage{graphicx}
\usepackage{epsfig}
\usepackage{bm}
\usepackage{amssymb}
\usepackage{float}
\usepackage{amsmath}
\usepackage{dcolumn}
\usepackage[normalem]{ulem}
\usepackage{cancel}
\usepackage[colorlinks]{hyperref}
\usepackage[usenames,dvipsnames]{color}
\hypersetup{
     breaklinks=true,
    pdfstartview={FitH},    
    colorlinks=true,       
    linkcolor=blue,          
    citecolor=red,        
    filecolor=magenta,      
    urlcolor=Green,           
    anchorcolor=blue,      
    linktocpage=true
}

\newcommand{\Mpl}{M_{\textrm{Pl}}}
\renewcommand{\(}{\left(}
\renewcommand{\)}{\right)}

\newcommand{\K}{\mathcal{K}}

\newcommand{\nn}{\nonumber}

\newcommand{\Eqn}[1]{&\hspace{-0.2em}#1\hspace{-0.2em}&}
\begin{document}

\title{Cosmological investigations of (extended) nonlinear massive gravity schemes with non-minimal coupling }

\author{K. Bamba} \email{bamba@kmi.nagoya-u.ac.jp}
\affiliation{Kobayashi-Maskawa Institute for the
Origin of Particles and the Universe, \\Nagoya University, Nagoya
464-8602, Japan}

\author{Md. Wali Hossain}\email{wali@ctp-jamia.res.in}
\affiliation{Centre for Theoretical Physics, Jamia Millia Islamia,
New Delhi-110025, India }

\author{R. Myrzakulov}\email{rmyrzakulov@gmail.com}
\affiliation{ Eurasian  International Center for Theoretical
Physics, Eurasian National University, Astana 010008, Kazakhstan}

\author{S. Nojiri}\email{nojiri@phys.nagoya-u.ac.jp}
\affiliation{Kobayashi-Maskawa Institute for the Origin of Particles
and the Universe, \\Nagoya University, Nagoya 464-8602, Japan}
\affiliation{Department of Physics, Nagoya University, Nagoya 464-8602, Japan}

\author{M. Sami}\email{sami@iucaa.ernet.in}
\affiliation{Centre for Theoretical Physics, Jamia Millia Islamia,
New Delhi-110025, India }

\begin{abstract}
In this paper we investigate the case of non-minimal coupling in
the (extended) nonlinear massive gravity theories. We first consider
massive gravity in the Brans-Dicke background such that the graviton
mass is replaced by $A^2(\sigma)m$ where $\sigma$ is the Brans-Dicke
field and $A(\sigma)$ is conformal coupling and show that there is
no viable thermal history of the universe in this case.  We then
invoke a cubic galileon term as nonlinear completion of the $\sigma$
Lagrangian and show that there is a stable de Sitter solution in this
case. However, the de Sitter is blocked by the matter phase which is
also a simultaneous attractor of the dynamics. The de Sitter phase
can, however, be realized by invoking unnatural fine tunings. We next
investigate cosmology of quasi-dilaton nonlinear massive gravity
with non-minimal coupling. As a generic feature of the non-minimal
coupling, we show that the model exhibits a transient phantom phase
which is otherwise impossible. While performing the observational
data analysis on the models, we find that a small value of coupling
constant is allowed for quasi-dilaton nonlinear massive gravity. For
both the cases under consideration, it is observed that we have an
effective pressure of matter which comes from the constraint
equation. For mass-varying nonlinear massive gravity in the Brans-Dicke
background, the effective pressure of matter is non zero which
affects the evolution of the Hubble parameter thereby spoiling
consistency of the model with data. As for, quasi-dilaton nonlinear
massive gravity, the effective pressure of matter can be kept around
zero by controlling the coupling constant, the model is shown to be fit
well with observations.
\end{abstract}


\maketitle

\section{Introduction}
There is an alternative thought in cosmology that late time cosmic
acceleration is not to due dark energy but is rather caused by the
large scale modification of Einstein theory of general relativity
(for review one can see \cite{Copeland:2006wr, Review-N-O, Clifton:2011jh, Bamba:2012cp}).
A modified theory of gravity should essentially reduce to the Einstein
theory along with some extra degree(s) of freedom. These degrees of
freedom directly couple with matter and might violate local physics
where the Einstein theory is an excellent agreement with observations
and thereby need to be screened out. There are two mechanisms of
hiding the extra degree(s) of freedom depending upon whether the
degrees of freedom are massive or massless. Chameleon screening
\cite{Khoury:2003aq,Khoury:2003rn}
relies on the fact that the mass of the field(s) representing the extra
degree(s) becomes environment dependent such that the field acquires
a large mass in high density regime escaping its detection locally.
In case of massless field(s), screening is achieved using the Vainshtein
mechanism \cite{Vainshtein:1972sx} which operates through kinetic
suppression which is a nonlinear phenomenon.

In chameleon supported theories, mass screening to the required
accuracy leaves no scope for large scale modification to be cause of
cosmic acceleration. The Vainshtein screening naturally arises in
galileon field theory which involves non linear completion of the
free scalar field. Galileons \cite{Luty:2003vm,Nicolis:2004qq,Nicolis:2008in},
in four dimensions, are the
reprehensive of the Lovelock structure \cite{VanAcoleyen:2011mj}
and thereby ghost free. They
occur in nonlinear massive gravity in the decoupling limit.

Massive gravity provides with a novel large scale modification that
could account for late time cosmic acceleration. Indeed, the
Newtonian potential for a static source with mass $M$ is given by
$GM \mathrm{e}^{-mr}/r$ in case the gravitational force is mediated by
graviton with mass $m$. The graviton mass should be typically of the
order of $H_0$ such that the modification is felt only at large
scales. The introduction of mass gives rise to the weakening of
gravity at large scales$-$ an effect which can be mimicked by
cosmological constant in the standard lore. It is certainly a novel
perspective that cosmological constant gets related to mass of
graviton in massive gravity.

A linear theory of massive gravity was formulated by Pauli and Fierz
in 1939 \cite{Fierz:1939ix} with a motivation to write down the
consistent relativistic equation for spin-2 field. The theory,
however, suffers from the van Dam-Veltman-Zakharov (vDVZ)
discontinuity problem \cite{vanDam:1970vg,Zakharov:1970cc}.
It was pointed out by
Vainshtein that the linear approximation is violated for a massive
body. The problem of discontinuity is solved putting the theory in
the non linear background but a ghost known as the Boulware-Deser (BD)
ghost kicks in \cite{Boulware:1973my}. This results sounds like a no
go theorem and it took many years to overcome the problem. The
nonlinear theory of massive gravity which is ghost free with the Vainshtein
screening in built is known as the de Rham, Gabadadze, Tolley ($dRGT$)
\cite{deRham:2010ik,deRham:2010kj} (for a review one can see
\cite{Hinterbichler:2011tt}). This theory is a consistent
generalization of Pauli-Fierz theory in nonlinear background. It
was shown by Hassan and Rosen \cite{Hassan:2011hr} that the
Hamiltonian constraint is maintained at the nonlinear order along
with the associated secondary constraint, which implies the absence
of the BD ghost.

It looks at the onset that we should be able to formulate the
Friedmann-Lema\^{i}tre-Robertson-Walker (FLRW)
cosmology in massive gravity. However, leaving the problem of
superluminality aside which an essential feature of any ghost free
nonlinear massive gravity \cite{Deser:2012qx},
cosmological solutions in the $dRGT$ theory have been examined
in Refs.~\cite{Koyama:2011xz, Koyama:2011yg}, and also it turns out that the
FLRW cosmology is absent in the $dRGT$ \cite{D'Amico:2011jj}: The
spatially flat geometry does not exist and the other two branches,
$K\pm 1$ are unstable \cite{DeFelice:2012mx},
where $K$ is the spatial curvature,
except the case that a particular frame is taken to
enhance the symmetry~\cite{Khosravi:2013axa}.
Let us note that the $dRGT$ admits inhomogeneous and anisotropic
 stable backgrounds \cite{D'Amico:2011jj,Gumrukcuoglu:2012aa,
 DeFelice:2013awa,Gumrukcuoglu:2011ew,Gumrukcuoglu:2011zh},
 and that this fact may be interesting in view of the recently
 observed large anisotropy in radio data(see \cite{Jain:1998kf} and references
 therein).

However, if we want to retain the standard framework, we need to
modify the $dRGT$ to reconcile with the homogeneity and isotropy.
Perhaps the simplest way out could be to replace the mass of
graviton by a field dependent quantity {\it a la} mass-varying
massive gravity
\cite{Huang:2012pe,Cai:2012ag,Huang:2013mha,Saridakis:2012jy,Leon:2013qh}
and quasi-dilaton nonlinear massive gravity
\cite{D'Amico:2012zv,Gannouji:2013rwa}. Though we have a self
accelerating solution in mass-varying massive gravity, the massive
gravity dominated era is an intermediate state and the late time
cosmic acceleration caused by the quintessence effect
\cite{Cai:2012ag,Saridakis:2012jy,Leon:2013qh}. On the other hand,
in the quasi-dilaton nonlinear massive gravity, the late time cosmic
acceleration comes from the massive part in the Lagrangian such that
the massive gravity dominated era is an attractor of the dynamics
\cite{D'Amico:2012zv,Gannouji:2013rwa}. In this setting, there is an
underlying symmetry and the graviton mass becomes a function of
dilaton, $\sigma$. The model has interesting cosmological features,
in particular, it has  de Sitter solution as an attractor of the
dynamics and a viable matter dominated regime. In this scenario,
things are arranged in such a way that $\sigma$ does not directly
couple to matter in the Einstein frame. However, the model suffers
from the ghost instability \cite{Gumrukcuoglu:2013nza,
D'Amico:2013kya} which signifies that the FRW background itself is
unstable. Efforts have recently been made to develop a consistent
extended quasi-dilaton massive gravity
\cite{DeFelice:2013tsa,DeFelice:2013dua} free from ghosts but with
the same background dynamics as in the original framework. Efforts
were also made to see the cosmology by coupling DBI galileon to the
massive gravity \cite{Andrews:2013ora,Andrews:2013uca}. Though this
model is ghost free \cite{Andrews:2013ora} it does not posses flat
FLRW solution \cite{Andrews:2013uca,Hinterbichler:2013dv}.
Cosmological perturbations in $dRGT$ and extended $dRGT$ have also
been seen in Refs. \cite{Gumrukcuoglu:2011zh,D'Amico:2013kya,
D'Amico:2012pi,Andrews:2013uca,Guarato:2013gba,Gumrukcuoglu:2013nza,Haghani:2013eya}.
Moreover, we mention that non-minimal couplings in massive gravity
have been investigated in the context of bimetric
gravity~\cite{Akrami:2013ffa}. This consideration has also been
generalized to multigravity in Ref.~\cite{Tamanini:2013xia}. Apart
from this, many works have been done on bimetric and multimetric
gravity
\cite{Hassan:2011zd,Volkov:2011an,vonStrauss:2011mq,Akrami:2012vf,Hassan:2011ea,
Khosravi:2011zi,Baccetti:2012bk,Berg:2012kn,Deser:2013gpa}.
Furthermore, tunneling effect between vacua~\cite{Zhang:2012ap} and
an application to quantum cosmology~\cite{Sasaki:2013nka} have been
studied.

In this paper, we investigate cosmological dynamics of extended
massive gravity models with non-minimal coupling.
We organize the structure of the paper as follows.
In Section \ref{mass-varying}, we study
cosmological behavior for different cases of mass-varying nonlinear
massive gravity in presence of non-minimal coupling. In sub sections
\ref{dlimit} $\&$ \ref{screening}, we discuss the decoupling limit
and the Vainshtein screening for the action (\ref{Eaction1}). Cosmology
based upon the action (\ref{Eaction1}) complimented with the cubic
galileon Lagrangian is discussed in sub section \ref{cosmology}. In
section \ref {quasidilaton}, we investigate cosmological dynamics of
non-minimally coupled quasi-dilaton nonlinear massive gravity.
Sub section \ref {data_analysis} is devoted to study of
observational constraints on the model parameters. Finally, in
section \ref{conclusions}, we summarize our results. In Appendix, we
present the method used to do the data analysis using the $\chi^2$
minimization method.

\section{Non minimal coupling and mass-varying massive gravity}
\label{mass-varying}

 Let us consider the following action in the Einstein frame
 \begin{align}
 \label{Eaction1}
 \mathcal{S}_E &=\int
d^4x\sqrt{-g}\Big[\frac{\Mpl^2}{2}R-\frac{1}{2}g^{\mu \nu}
\partial_\mu \sigma \partial_\nu
\sigma-V(\sigma)\nn \\ &-\frac{m^2\Mpl^2}{4}A^{4}\left(4\,\mathcal{U}_2(\mathcal{K})+\alpha_3\mathcal{U}_3(\mathcal{K})+
\alpha_4\mathcal{U}_4(\mathcal{K})\right)\Big]\nn \\
&+\int d^4x\sqrt{-g}\mathcal{L}_m(A^2g_{\mu \nu},\psi)\,,
 \end{align}
where $\tilde{g}_{\mu\nu}$ is the Jordan
frame metric related to the Einstein frame metric $g_{\mu\nu}$ by the relation
\begin{equation}
 \tilde{g}_{\mu\nu}=A^2(\sigma)g_{\mu \nu}\,.
\end{equation}
Here $A^2(\sigma)$ is the conformal factor relating the two metrics of the two frames. For simplicity we assume
the coupling $\beta=-\Mpl ~ ({\rm d}\ln A/{\rm d}\sigma)$ to be constant.
This leads to
\begin{equation}
A(\sigma)=\mathrm{e}^{-\beta\sigma/\Mpl}\,, \label{conffactor}
\end{equation}
where $\Mpl$ is the reduced Planck mass. In the
action (\ref{Eaction1}) $m$ is the graviton mass, $\sigma$ is a
scalar field and $V(\sigma)$ is the potential of the $\sigma$ field.
In the mass term we have a coupling between the dilaton field and
the massive gravity. Due to the conformal factor the mass of the
graviton $m$ in pure $dRGT$ is replaced here by the term $mA^2(\sigma)$ giving
rise to a mass-varying nonlinear massive gravity action which can possibly give
a viable cosmology.

The functions $\mathcal{U}_i$ $(i=2,3,4)$ are given by
\begin{widetext}
\begin{subequations}
\begin{eqnarray}
&&\mathcal{U}_2(\K)=\frac{1}{2!}\varepsilon_{\mu\alpha ..}\varepsilon^{\nu \beta ..}\K_\nu^\mu \K_\beta^\alpha ~=
[\K^2]-[\K]^2 \,, \\
&&\mathcal{U}_3(\K)=\frac{1}{1!}\varepsilon_{\mu\alpha\gamma .}\varepsilon^{\nu\beta\delta .} 
\K^\mu_\nu\K^\alpha_\beta\K^\gamma_\delta ~=
-[\K]^3+3[\K][\K^2]-2[K^3] \,, \\
&&\mathcal{U}_4(\K)=\frac{1}{0!}\varepsilon_{\mu\alpha\gamma \rho}\varepsilon^{\nu\beta\delta\sigma}\K^\mu_\nu\K^\alpha_\beta
\K^\gamma_\delta \K^\rho_\sigma ~= -[\K]^4+6[\K^2][\K]^2-8[\K^3][\K]-3[\K^2]^2+6[\K^4]\,,
\end{eqnarray}
\end{subequations}
\end{widetext}
where $[\K]=g^{\mu\nu}\K_{\mu\nu}$, $[\K^2]=\K^\mu_\nu\K^\nu_\mu$, $[\K^3]=\K^\mu_\nu\K^\nu_\alpha\K^\alpha_\mu$ and
$[\K^4]=\K^\mu_\nu\K^\nu_\alpha\K^\alpha_\beta\K^\beta_\mu$. The matrix $\mathcal{K}^\mu_\nu$ has the following form
\begin{equation}
\mathcal{K}^\mu_\nu=\delta^\mu_\nu-\sqrt{g^{\mu\alpha}\partial_\alpha\phi^a\partial_\nu\phi^b \eta_{ab}}
\end{equation}
where $a,b\in (0,1,2,3)$ are internal indices,
$\eta_{ab}=\text{diag}(-1,1,1,1)$ and $\phi^a$ are the four
St\"{u}ckelberg scalars introduced to restore general covariance
\cite{ArkaniHamed:2002sp}.

It would be instructive to cast the action (\ref{Eaction1}) in the
Jordan frame
\begin{eqnarray}
\label{Eaction2}
&&\mathcal{S}_J =\int d^4x\sqrt{-\tilde{g}}\Big[\frac{\Mpl^2}{2}\tilde{R}\Psi \nn \\ && {}-\frac{\Mpl^2}{2}
\frac{\omega_{\rm BD}(\Psi)}{\Psi}\tilde{g}^{\mu\nu}\left(\partial_\mu \Psi
\partial_\nu
\Psi\right) -\Psi^2 V(\Psi) \nn \\ &&
{}-\frac{m^2\Mpl^2}{4}\left(4\,\mathcal{U}_2(\tilde{\mathcal{K}})+\alpha_3\mathcal{U}_3(\tilde{\mathcal{K}})+
\alpha_4\mathcal{U}_4(\tilde{\mathcal{K}})\right)\Big],
\end{eqnarray}
where the quantities with the $tilde$ represent the corresponding
Einstein frame quantities in the Jordan frame with $\omega_{\rm BD}$
as the Brans-Dicke parameter. The $\tilde{\mathcal{K}}^\mu_\nu$
matrices are given by,
\begin{equation}
\tilde{\mathcal{K}}^\mu_\nu=\delta^\mu_\nu-A(\sigma)\sqrt{\tilde{g}^{\mu\alpha}\partial_\alpha\phi^a\partial_\nu\phi^b\eta_{ab}}
~;~~ \text{and}~~\Psi\equiv A^{-2}\,,
\end{equation}
which reduces to $\K^\mu_\nu$ after using the conformal
transformation. It is observed that if we choose, $\beta=-1$, the
action (\ref{Eaction2}) becomes same as the one in quasi-dilaton
nonlinear massive gravity \cite{D'Amico:2012zv} in the Jordan frame
without the potential $V(\sigma)$. Now to make the coefficient in
front of the kinetic term of the $\sigma$ field in the Einstein
frame $1/2$ with a right sign,
we need to have \cite{DeFelice:2010aj,Fujii_Maeda}
\begin{equation}
 \omega_{\rm BD}=\frac{1}{2}\left[\frac{1}{2\Mpl^2 (d\ln A/d\sigma)^2}
-3\right]\,,
\end{equation}
the conformal factor(\ref{conffactor}) in case of which gives
\begin{equation}
\omega_{\rm BD}=\frac{1-6\beta^2}{4\beta^2}\,.
\end{equation}

Let us for simplicity omit the potential and notice that if we
choose $\beta=1/\sqrt{6}$, kinetic term for $\sigma$
 field would be absent in Jordan frame but would reappear
in the Einstein frame by virtue of conformal transformations.
 Then the Jordan frame action would be simplified to
 \begin{align}
\label{Eaction4}
\mathcal{S}_J &=\int d^4x\sqrt{-\tilde{g}}\Big[\frac{\Mpl^2}{2}\tilde{R}e^{2\beta\sigma/\Mpl}-
\frac{m^2\Mpl^2}{4}\left(4\,\mathcal{U}_2(\tilde{\mathcal{K}}) \right. \nn\\
& \left.+\alpha_3\mathcal{U}_3(\tilde{\mathcal{K}})+
\alpha_4\mathcal{U}_4(\tilde{\mathcal{K}})\right)\Big]
+\int d^4x\sqrt{-\tilde{g}}\mathcal{L}_m(\tilde{g}_{\mu\nu},\psi)
\end{align}
The latter is nothing but the massive gravity action in the
Brans-Dicke background. However, we shall consider the generalized
Brans-Dicke action (\ref{Eaction2}). We should go to the Einstein
frame to diagonalize the Jordan frame action at the cost of coupling to
matter and obtain (\ref{Eaction1}). However, we replace the
graviton mass with a function of the $\sigma$ field thereby allowing us to
have the FLRW cosmology, which is otherwise not possible to be realized.
The viability of cosmology should then be checked.

\subsection{Decoupling limit and the Vainshtein screening}
\label{dlimit}

The decoupling limit is a sort of high
energy limit such that the energy scales are much larger than the
graviton mass
\begin{align}
& \Mpl\to \infty\,,~~~m \to 0\,,~~~\Lambda_3=(\Mpl
m^2)^{1/3}=\text{fixed}\,,~~~ \nn \\ &\frac{T_{\mu\nu}}{\Mpl}=\rm fixed\,.
\label{dcplng}
\end{align}
Here, we mention that the ways of taking a decoupling limit have also been 
investigated in Refs.~\cite{deRham:2010tw, deRham:2011by}.

Secondly, we shall canonically normalize the fields: $h_{\mu\nu}\to
2\hat{h}_{\mu\nu}/\Mpl,\pi \to \hat{\pi}/m^2\Mpl$ and ignore the
helicity 1 field, they get decoupled from matter in this limit. One
defines the covariant tensor $H_{\mu\nu}$ as
$H_{\mu\nu}=g_{\mu\nu}-\partial_\mu
\phi^a\partial_\nu\phi^b\eta_{ab}$. {}From this expression, we find
\begin{align}
 g_{\mu\nu}=\eta_{\mu\nu}+\frac{2\hat{h}_{\mu\nu}}{\Mpl}=H_{\mu\nu}+\partial_\mu \phi^a\partial_\nu\phi^b\eta_{ab}.
\end{align}
The canonically normalized helicity-0 graviton $\pi$ is defined as
$\phi^a=\delta^a_\mu x^\mu-\eta^{a\mu}\partial_\mu \hat{\pi}/\Lambda_3^3$, which yields
\begin{align}
 H_{\mu\nu}=\frac{2\hat{h}_{\mu\nu}}{\Mpl}+\frac{2\Pi_{\mu\nu}}{\Lambda_3^3}-\frac{\Pi_{\mu\nu}^2}{\Lambda_3^6}\,,
\end{align}
where $\Pi_{\mu\nu}=\partial_\mu\partial_\nu\hat{\pi}$,
$\Pi_{\mu\nu}^2=\Pi_{\mu\alpha}\Pi^\alpha_{\nu}$ and
$\Pi=\Pi^\mu_\mu$.

In the decoupling limit, only the linearized term survives from the
Einstein-Hilbert action; the $\sigma$ field kinetic term remains
unaffected. The total derivative terms do not disappear from the
description because they now figure with $\sigma$ and we have the first
order interactions containing the galileon terms. In the action
(\ref{Eaction1}), in decoupling limit, we imagine $\mathrm{e}^{-4\beta
\sigma/\Mpl}\simeq 1-4\beta \sigma/\Mpl$ in the mass term. The first
term in decoupling limit would provide the interaction term plus the
total derivative. The latter has no effect on dynamics. However, the
total derivative term multiplied by $\sigma$ (coming from the second
term of exponential expansion) has dynamics. Thus, we obtain
\begin{align}
\label{dcL2}
 &\mathcal{L}_{dc} =\frac{1}{2}\hat{h}^{\mu\nu}(\mathcal{E}\hat{h})_{\mu\nu}-\frac{1}{2}\partial^\mu\sigma\partial_\mu
\sigma\nn \\ &+2\hat{h}^{\mu\nu}\Lambda_3^3\left[2X^{(1)}_{\mu\nu}+\frac{(3\alpha_3+4)}{4}
X_{\mu\nu}^{(2)}+\frac{(\alpha_3+4\alpha_4)
}{4}X_{\mu\nu}^{(3)}\right]\nn \\ &
+\beta\sigma\left[4\frac{\mathcal{U}_2(\Pi)}{\Lambda^3_3}+\alpha_3\frac{\mathcal{U}_3(\Pi)}{\Lambda^3_6}+
\alpha_4\frac{\mathcal{U}_4(\Pi)}{\Lambda^9_3}\right] \nn \\ & +\frac{2}{\Mpl}\hat{h}_{\mu\nu}T^{\mu\nu}\,,
\end{align}
where the action of the Einstein operator on a symmetric object
$Z_{\alpha\beta}$ is given by
\begin{align}
(\mathcal{E}Z)_{\mu\nu} =& \Box
Z_{\mu\nu}-\eta_{\mu\nu}\Box Z-\partial_\mu \partial^\alpha
Z_{\nu\alpha}-\partial_\nu \partial^\alpha
Z_{\mu\alpha} \nn \\ & +\partial_\mu\partial_\nu Z+\eta_{\mu\nu}\partial^\alpha\partial^\beta
Z_{\alpha\beta}\equiv\mathcal{E}^{\alpha\beta}_{\mu\nu}Z_{\alpha\beta}\,.
\end{align}
Here we have used (following \cite{deRham:2010ik,Hinterbichler:2011tt})
\begin{eqnarray}
&& \frac{\delta}{\delta h^{\mu
\nu}}\left(\sqrt{-g}\mathcal{U}_n\right)|_{h_{\mu\nu}=0} = \bar{X}^{\mu\nu}_{(n)} \nn \\ && \equiv
\Lambda_3^3 \sum_{n\geq2}\alpha_n\left(X^{(n)}_{\mu\nu}+nX^{(n-1)}_{\mu\nu}\right)\,,
\end{eqnarray}
where
\begin{equation}
X^{(n)}_{\mu\nu}=\frac{1}{\Lambda_3^3}\Big[-n\Pi^\alpha_\mu X^{(n-1)}_{\alpha\nu}+
\Pi^{\alpha\beta}X^{(n-1)}_{\alpha\beta}\eta_{\mu\nu}\Big]\,,
\end{equation}
which gives
\begin{subequations}
\begin{eqnarray}
&&X_{\mu\nu}^{(1)}=-\frac{1}{4\Lambda_3^3}\varepsilon_{\mu\alpha}^{~~~~..}\varepsilon_{\nu\beta ..}
\Pi^{\alpha\beta}\,, \\
&&X_{\mu\nu}^{(2)}=-\frac{1}{2\Lambda_3^6}\varepsilon_{\mu\alpha\rho}^{~~~~~.}\varepsilon_{\nu\beta\sigma .}
\Pi^{\alpha\sigma}\Pi^{\beta \rho} \,, \\
&&X_{\mu\nu}^{(3)}=-\frac{1}{2\Lambda_3^9}\varepsilon_{\mu\alpha \rho\gamma}\varepsilon_{\nu\beta \sigma\delta}
 \Pi^{\alpha\sigma}\Pi^{\beta\gamma}\Pi^{\rho\delta}\,,
\end{eqnarray}
\end{subequations}
with $X^{(0)}_{\mu\nu}=1/2~\eta_{\mu\nu}$. Here we have also used the fact that $X^{(n)}_{\mu\nu}=0$ for all $n\geq 4$.
We notice that the next
term in the expansion containing $\sigma/\Mpl$ would vanish in the
decoupling limit. Also notice that the factor $\Mpl^2/4$, which one gets
expanding the Einstein-Hilbert action to the quadratic level, is taken
care of by the field redefinition in the first term of the action.

The action (\ref{dcL2}) can be partially diagonalized through a
linear conformal transformation. Since
$\hat{h}^{\mu\nu}X_{\mu\nu}^{(3)}$ cannot be diagonalized by any
local transformation, we remove this term by choosing its coefficient
to be equal to zero as $\alpha_3=-4\alpha_4$. As for the conformal
transformation
\begin{equation}
\hat{h}_{\mu\nu}\to \hat{h}_{\mu\nu}+2\hat{\pi}\eta_{\mu\nu}+2\frac{D}{\Lambda^3_3}\partial_\mu \hat{\pi}\partial_\nu
\hat{\pi}\,;~D=-(1-3\alpha_4)\,,
\end{equation}
that we make on the action (\ref{dcL2}) in sequence, we notice the
following
\begin{subequations}
\begin{eqnarray}
&&\mathcal{E}^{\alpha\beta}_{\mu\nu}Z_{\alpha\beta}=-4\Lambda_3^3 X^{(1)}_{\mu\nu}\,,
~Z_{\alpha\beta}=\hat{\pi}\eta_{\alpha\beta}\,, \\
&&\mathcal{E}^{\alpha\beta}_{\mu\nu}Z_{\alpha\beta}=2\Lambda_3^6 X^{(2)}_{\mu\nu}\,,~
Z_{\alpha\beta}=\partial_\alpha\hat{\pi}\partial_\beta \hat{\pi}\,, \\
&&\hat{\pi} \eta^{\mu\nu}X^{(1)}_{\mu\nu}=\frac{3}{2\Lambda_3^3}\hat{\pi}\Box\hat{\pi}
=-\frac{1}{4\Lambda_3^3}\varepsilon\varepsilon \Pi\,.
\end{eqnarray}
\end{subequations}
The decoupling limit Lagrangian finally acquires the form
\begin{widetext}
\begin{eqnarray}
\label{dcL3}
\hspace{-3mm}
 &&\mathcal{L}_{dc}=\frac{1}{2}\hat{h}^{\mu\nu}(\mathcal{E}\hat{h})_{\mu\nu}-\frac{1}{2}\partial^\mu\sigma\partial_\mu
\sigma-2\hat{\pi}\varepsilon\varepsilon\Pi+\frac{4D}{\Lambda_3^3}\hat{\pi}\varepsilon\varepsilon \Pi\Pi
-\frac {2D^2 }{\Lambda^6_6}\hat{\pi} \varepsilon\varepsilon \Pi\Pi\Pi
+\beta\sigma\left[\frac{2}{\Lambda^3_3}\varepsilon\varepsilon\Pi\Pi-\frac{4\alpha_4}{\Lambda^3_6}\varepsilon\varepsilon\Pi\Pi\Pi+
\frac{\alpha_4}{\Lambda^9_3}\varepsilon\varepsilon\Pi\Pi\Pi\Pi\right] \nn \\
\hspace{-3mm}
&& +\frac{2}{\Mpl}\hat{h}_{\mu\nu}T^{\mu\nu}+\frac{2}{\Mpl}\hat{\pi}
T +\frac{1}{\Mpl}\frac{2D}{\Lambda^3_3}\partial_\mu
\hat{\pi}\partial_\nu \hat{\pi}  T^{\mu\nu}\,,
\end{eqnarray}
\end{widetext}
where
$\varepsilon\varepsilon\Pi=\varepsilon_{\mu...}\varepsilon_\nu^{~~...}\Pi^{\mu\nu}$,
$\varepsilon\varepsilon\Pi\Pi=\varepsilon_{\mu\alpha..}\varepsilon_{\nu\beta}^{~~~..}\Pi^{\mu\nu}\Pi^{\alpha\beta}$,
$\varepsilon\varepsilon\Pi\Pi\Pi=\varepsilon_{\mu\alpha\gamma
.}\varepsilon_{\nu\beta\rho}^{~~~~~.}
\Pi^{\mu\nu}\Pi^{\alpha\beta}\Pi^{\gamma\rho}$ and
$\varepsilon\varepsilon\Pi\Pi\Pi\Pi=\varepsilon_{\mu\alpha\gamma\lambda}\varepsilon_{\nu\beta\rho\delta}
\Pi^{\mu\nu}\Pi^{\alpha\beta}\Pi^{\gamma\rho}\Pi^{\lambda\delta}$
\cite{D'Amico:2012zv}. Here we have also used
$\eta^{\mu\nu}\varepsilon_\mu\varepsilon_\nu\Pi\Pi=
\Pi^{\mu\nu}\varepsilon_\mu\varepsilon_\nu\Pi$ by keeping in mind
that $\partial^{\mu}X^{(n)}_{\mu\nu}=0$ and
$\partial^{\mu}\Pi^{(n)}_{\mu\nu}=0$
\cite{Mirbabayi:2011aa,Hinterbichler:2011tt}.

Let us notice that all the terms appearing in the above Lagrangian
(\ref{dcL3}) belong to the class of galileons. For instance, in the terms
containing the product of $\sigma$ and $\mathcal{U}^{'s}$, one of
the differential operators from $\partial\partial \pi$ can be
shifted to $\sigma$ by means of integration by parts thereby
converting $\sigma \mathcal{U}^{'s}$ into a corresponding galileon
Lagrangian.

\subsubsection{Screening and local physics}
\label{screening}

The decoupling Lagrangian  provides  an accurate description of
local physics {\it a la} the Vainshtein affect. It is interesting to
note that the direct coupling of $\sigma$ to the metric completely
disappears in the decoupling limit; the longitudinal mode of
graviton $\hat{\pi}$ alone directly couples to gravity. Hence
$\sigma$ has no direct impact locally though $\sigma$ is coupled to
$\hat{\pi}$. However, the non minimal coupling can have impact on
large scales.
 As for the local physics,  we have to worry about
the screening of $\hat{\pi}$ and in what follows we check the same.

The equations of motion for $\sigma$ and $\hat{\pi}$ which follow
from the Eq.~(\ref{dcL3}) have the following forms
\begin{eqnarray}
&&\Box\sigma+\frac{2\beta}{\Lambda_3^3}\varepsilon\varepsilon\Pi\Pi-4\frac{\alpha_4\beta}{\Lambda_3^6}\varepsilon\varepsilon
\Pi\Pi\Pi+\frac{\beta\alpha_4}{\Lambda_3^9}\varepsilon\varepsilon\Pi\Pi\Pi\Pi  \nn \\ && =0 \,, \\
&&-4\varepsilon\varepsilon\Pi+\frac{12D}{\Lambda_3^3}\varepsilon\varepsilon\Pi\Pi-\frac{8D^2}{\Lambda_3^6}\varepsilon\varepsilon
\Pi\Pi\Pi+\frac{4\beta}{\Lambda_3^3}\varepsilon\varepsilon\Sigma\Pi- \nn \\ && \frac{12\beta\alpha_4}{\Lambda_3^6}
\varepsilon\varepsilon\Sigma\Pi\Pi+\frac{4\beta\alpha_4}{\Lambda_3^9}\varepsilon\varepsilon\Sigma\Pi\Pi\Pi=-\frac{2}{M_P}T\,,
\label{vpi}
\end{eqnarray}
where $\Sigma=\partial_\mu\partial^\mu\sigma$. Since in this model
the Einstein frame metric is coupled to the $\sigma$ field through
conformal coupling, the matter conservation equation gets modified
here and it becomes $\partial_\mu
T^\mu_\nu=\frac{\beta}{\Mpl}T\partial_\nu\sigma$ (here we have used
partial derivative because we are working in the decoupling limit).
Hence it is possible to think that we can have another term in Eq.~(\ref{vpi})
coming from the last term of the Eq.~(\ref{dcL3}) due to the modified
matter conservation equation. However, in the modified matter
conservation equation we have a term with the factor $\beta/\Mpl$,
thanks to which the contribution from the modified matter
conservation equation will vanish in the decoupling limit. Thus we
do not have any contribution in Eq.~(\ref{vpi}) from the
conformal transformation. Here we should note that again we have
used the property of $X^{(n)}_{\mu\nu}$ that $\partial^\mu
X^{(n)}_{\mu\nu}=0$.

For a static spherically symmetric source, $T_{00}=-M\delta(r)/r^2$, the
above equations can be integrated and eventually give rise to algebraic equations for $\hat{\pi}'/r$ and $\sigma'/r$ as
\begin{eqnarray}
\label{lsigma}
&&\lambda_\sigma-8\beta\(\lambda_\pi^2-\alpha_4\lambda_\pi^3\)=0\,,    \\
\label{lpi}
&& 12\lambda_\pi-24 D\lambda_\pi^2+8 D^2\lambda_\pi^3-8\beta\lambda_\pi\lambda_\sigma+
12\beta\alpha_4\lambda_\pi^2\lambda_\sigma \nn \\ && =\left(\frac{r_V}{r}\right)^3 \,,
\end{eqnarray}
where $\lambda_\pi=\hat{\pi}'/\Lambda_3^3r$, $\lambda_\sigma=\sigma'/\Lambda_3^3r$, and
$r_V=\left(M/(m^2\Mpl^2)\right)^{1/3}$ is the Vainshtein radius.

Putting the value of $\lambda_\sigma$ from Eq.~(\ref{lsigma}) to
Eq.~(\ref{lpi}), we can have
\begin{eqnarray}
\label{lpi1}
&& 12\lambda_\pi -24 D\lambda_\pi^2+8\(D^2-8\beta^2\)\lambda_\pi^3+160\alpha_4 \beta^2 \lambda_\pi^4-\nn \\ &&
96\alpha_4^2\beta^2\lambda_\pi^5  =\left(\frac{r_V}{r}\right)^3 \,.
\end{eqnarray}

At distances much smaller than the Vainshtein radius ($r\ll r_v$), keeping the
highest non-linearity, we get
\begin{eqnarray}
&& \lambda_\pi\simeq
-\left(\frac{r_V}{r}\right)^{3/5}\left(\frac{1}{96\beta^2\alpha_4^2}\right)^{1/5}\,,  \\ && \Rightarrow
\frac{\pi}{h_{00}}\propto \left(\frac{r}{r_V}\right)^{12/5}\,,
\end{eqnarray}
where $h_{00}$ is the gravitational potential which goes as $\sim
1/r$ and $\pi$ behaves as the potential for the fifth force.
Therefore from the last equation we can see that for $r\ll r_V$,
the gravitational potential dominates over the potential responsible for
the fifth force. Thus local physics is restored here through the Vainshtein
mechanism \cite{Vainshtein:1972sx} and we naturally get rid of the
vDVZ discontinuity \cite{Zakharov:1970cc,vanDam:1970vg}.

On the other hand, for $r\gg r_V$, the nonlinear effect gets
suppressed and the dominating term in Eq.~(\ref{lpi1}) becomes
$\lambda_\pi$. {}From Eq.~(\ref{lpi1}), we find that the fifth
force $\pi' \sim r_V^3/r^2$. This means that beyond the Vainshtein
radius the fifth force behaves like the Newtonian gravitational
force.

We also notice that at distances much smaller than $r_V$,
the relation
$\lambda_\sigma \simeq \lambda^3_\pi$ informs us that the screening of
$\sigma$ is much better than that of $\pi$.

\subsection{Cosmology}
\label{cosmology}

The model under consideration (action (\ref{Eaction1})) is similar
to the mass-varying nonlinear massive gravity models considered in
Refs.~\cite{Huang:2012pe,Saridakis:2012jy,Leon:2013qh,Cai:2012ag,Hinterbichler:2013dv},
where it has been shown that a non trivial flat FLRW solution
can be achieved.
{In the model under consideration, mass varying
term is due to the conformal coupling of matter with the field
$\sigma$ in the Einstein frame whereas in the original mass-varying
nonlinear massive gravity \cite{Huang:2012pe}, mass-varying term is
related to the potential term which is replaced with the mass squared in
the Lagrangian. One should also notice that in case of the action
(\ref{Eaction1}), the limit $\beta\to 0$ brings us to the original
$dRGT$ nonlinear massive gravity, which has the problem of not having
any non-trivial flat universe solution. In our case also the limit
$\beta\to 0$ gives the same problem. Here we will study the effects
of the non minimal coupling in models of mass-varying nonlinear
massive gravity.  We would first investigate the cosmological
dynamics for the system based upon the action (\ref{Eaction1})
without the potential term. Though it has been argued in
Refs.~\cite{Leon:2013qh,Cai:2012ag,Saridakis:2012jy} that one
needs an extra potential term other than the potential which
represents the graviton mass squares term to have late time
acceleration. In that case, the de Sitter solution results from the
extra potential term rather than the mass of the graviton. However, in
the model under consideration, we have conformal coupling factor
which modifies the matter continuity equation. It would therefore be
interesting to check whether the non minimal coupling can change
the situation. Next we will check the impact of adding the cubic
galileon Lagrangian ($L_3=-\frac{1}{2}
\frac{c_3}{M^3}(\nabla\sigma)^2 \Box\sigma$) in the
action (\ref{Eaction1}) without the potential. We should emphasize
that in galileon theory, the cubic order Lagrangian cannot give
rise to the self accelerating solution in the standard framework,
and that one needs at least the fourth order galileon to execute the
task~\cite{Gannouji:2010au,Ali:2010gr}. Adding a potential to the cubic
order galileon Lagrangian may also give rise to self acceleration
\cite{Ali:2012cv}. Finally we will study the effect of the non
minimal coupling in cosmology analyzing the action (\ref{Eaction1}).
In what follows we set up the necessary equations
considering the full action (\ref{Eaction1}) along with the cubic galileon
action in the presence of matter and radiation.

Let us suppose the spatially flat FLRW background of the form
\begin{equation}
 ds^2 = g_{\mu\nu}dx^\mu dx^\nu =  -N(t)^2 dt^2+a(t)^2 \delta_{ij}dx^i dx^j
\,,
 \label{metric}
\end{equation}
with $N(t)$ a function of $t$,
$\phi^0=f(t)$ and $\phi^i=a_{\rm ref}~x^i$, where $a_{\rm ref}$ is a reference
scale factor which can be set at very small values
like $\lesssim 10^{-9}$ to preserve the thermal history of the universe
\cite{Leon:2013qh}. Here in the choice of the St\"{u}ckelberg field we
have taken an arbitrary constant $a_{\rm ref}$ to avoid the inconsistency that we obtain at
 $a=1$ by choosing $\phi^i=x^i$. Generally this type of an arbitrary choice of
the St\"{u}ckelberg field is not possible \cite{Hinterbichler:2013dv} but as
argued in Ref.~\cite{Leon:2013qh} we can take this ansatz for the St\"{u}ckelberg
field as long as our fiducial metric is the Minkowskian.

The spatial volume in action (\ref{Eaction1}) can be integrated out from
the measure and we can rewrite the action as
\begin{eqnarray}
\mathcal{S} &=& \frac{\Mpl^2}{2} \int dt ~ \Big[- 6 \frac{a}{N}\dot{a}^2+
\frac{1}{\Mpl^2} \frac{a^3}{N} \dot{\sigma}^2 + \nn \\ &&
3 m^2 \mathrm{e}^{-4\beta\sigma/\Mpl} a^3\Big( NF_1(\xi)-\dot{f}F_2(\xi)\Big) \Big] - a^3 N V(\sigma)\nn \\ &&
-\frac{c_3}{M^3}\int \frac{a^2\dot{a}\dot{\sigma}^3}{N^3}dt+\mathcal{S}_m+ \mathcal{S}_r \,,
\label{Eaction5}
\end{eqnarray}
where $\mathcal{S}_m$ and $\mathcal{S}_r$ are the matter and the
radiation actions, respectively. Here matter is coupled with
the $\sigma$ field which modifies the continuity equation of matter.

Varying the action (\ref{Eaction5}) with respect to $N$ and setting $N=1$
at the end, i.e, $\delta \mathcal{S}/\delta N|_{N=1}=0$,
we find the Friedmann equation
\begin{eqnarray}
  3\Mpl^2 H^2 =\frac{\dot{\sigma}^2}{2}+\rho_m+\rho_r+\rho_{\rm mg}+\rho_{\rm gal}+V(\sigma)\,.
  \label{frd1}
  \end{eqnarray}

By imposing the condition $\delta \mathcal{S}/\delta a|_{N=1}=0$, we acquire
another gravitational field equation
\begin{equation}
   \(2\dot{H}+3H^2\)\Mpl^2 =-\frac{1}{2}\dot{\sigma}^2-p_{\rm mg}-\frac{1}{3}\rho_r-p_{\rm gal}+V(\sigma)\,,
  \label{frd2}
\end{equation}
where $\rho_{\rm mg}$, $\rho_{\rm gal}$, $p_{\rm mg}$, and $p_{\rm
gal}$ are the energy densities and pressures coming from the massive
gravity part and the cubic galileon Lagrangian, respectively, given by
\begin{eqnarray}
 \rho_{\rm mg}\Eqn{=}-\frac{3}{2}m^2\Mpl^2F_1(\xi)\mathrm{e}^{-4\beta\sigma/\Mpl}\,,
 \label{rho_l} \\
 p_{\rm mg} \Eqn{=} m^2\Mpl^2\mathrm{e}^{-4\beta\sigma/\Mpl}\Big(F_3(\xi)+\frac{1}{2}\dot{f}F_1'(\xi)\Big)\,,
\label{p_l}\\
\rho_{\rm gal} \Eqn{=} -\frac{3c_3}{M^3}H\dot{\sigma}^3 \,,
 \label{rho_gal}\\
 p_{\rm gal} \Eqn{=} \frac{c_3}{M^3}\dot{\sigma}^2\ddot{\sigma}\,,
 \label{p_gal}
\end{eqnarray}

where we have defined the following quantities for convenience
\begin{subequations}
\begin{eqnarray}
\hspace{-5mm}
\xi \Eqn{=} \frac{a_\mathrm{ref}}{a}\,, \\
\hspace{-5mm}
F_1(\xi) \Eqn{=} (1-\xi) \left[ 4(2-\xi)+\alpha_3(1-\xi)(4-\xi) \right. \nn \\
\hspace{-5mm}
&& \left. +4\alpha_4 (1-\xi)^2 \right] \,,  \\
\hspace{-5mm}
F_2(\xi) \Eqn{=} (1-\xi)\left[4+3\alpha_3(1-\xi)+4\alpha_4(1-\xi)^2\right]\,, \\\hspace{-5mm}
F_3(\xi) \Eqn{=} 2\left[6(1-\xi)+\xi^2\right]+3\alpha_3(1-\xi)(2-\xi) \nn \\
\hspace{-5mm}
&&  +6\alpha_4(1-\xi)^2 \,,
\end{eqnarray}
\end{subequations}
and the $\prime$ represents the derivative with respect to (w.r.t.) $\xi$.

The variation of the action (\ref{Eaction5}) with respect to $f(t)$ yields
a constraint equation
\begin{equation}
\mathrm{e}^{-4\beta\sigma/\Mpl}F_2(\xi)=\frac{C}{a^3}\,,
\label{constraint}
\end{equation}
where $C$ is an integration constant.

As for the field equation for $\sigma$, it has the following form
\begin{eqnarray}
&& \ddot{\sigma}+3H\dot{\sigma}+6\beta m^2\Mpl \mathrm{e}^{-4\beta\sigma/\Mpl}\(F_1(\xi)-\dot{f}F_2(\xi)\)  \nn \\ &&
-\frac{3c_3}{M^3} \dot{\sigma}\Bigl(3H^2\dot{\sigma}+\dot{H}\dot{\sigma}+2H\ddot{\sigma}\Bigr) +V'(\sigma)
\nn \\ &&
=-\frac{\beta}{\Mpl}\rho_m \,.
\label{sigeqn}
\end{eqnarray}

Now if we consider $C\neq 0$, then from Eq.~(\ref{constraint}), we
have the expression for $\dot{\sigma}$ and hence we can define a
density parameter for the kinetic part of the $\sigma$ field as
\begin{equation}
 \Omega_\sigma=\frac{1}{96 \beta^2}\(\frac{F_1'}{F_2}\)^2 \,.
 \label{Omsig}
\end{equation}
We can also define a density parameter for the massive gravity part
and the cubic galileon term as follows
\begin{eqnarray}
 \Omega_{\rm mg} \Eqn{=} -\frac{1}{2}\frac{m^2}{H^2}\frac{C}{a^3}\frac{F_1(\xi)}{F_2(\xi)} \,,
 \label{Ommg}\\
  \Omega_{\rm gal} \Eqn{=} \frac{c_3 \Mpl}{4^3 \beta^3
  M^3}H^2\(\frac{F_1'}{F_2}\)^3 \, .
 \label{Omgal}
\end{eqnarray}

Continuity equations of matter and radiation are given by
\begin{eqnarray}
  \dot{\rho_m}+3H\rho_m \Eqn{=} \frac{\beta}{\Mpl}\dot{\sigma}\rho_m \,,
 \label{contmat}\\
  \dot{\rho_r}+4H\rho_r \Eqn{=} 0 \,.
 \label{contrad}
\end{eqnarray}
Equation (\ref{contmat}) informs us that the evolution of matter density is
modified by virtue of its interaction with other components.
 Using Eqs.~(\ref{constraint}) and (\ref{contmat}), we have the following relation
\begin{equation}
\dot{\rho_m}+3H\rho_m\(1+\frac{F_1'}{12F_2}\)=0 \,,
\end{equation}
which implies that matter acquires an effective pressure
$p_{m(\text{eff})}=\rho_m\frac{F_1'}{12F_2}$ with an effective
equation of state parameter $w_{m(\text{eff})}=\frac{F_1'}{12F_2}$.
Hence due to the presence of conformal coupling, matter phase does
not have the conventional equation of state; it has an effective
non zero pressure.

\begin{table*}[ht]
\begin{center}
\caption[crit]{The critical points of the autonomous system
(\ref{eqxi1})-(\ref{eqr1}),  their existence conditions, the
eigenvalues of the perturbation matrix, and the  stability
conditions are given here. The numerical values of effective
equation of state ($w_{\text{eff}}$) and dark energy equation of
state ($w_{\text{DE}}$) are also included.} \label{tab1}
\resizebox{\textwidth}{!}{%
\begin{tabular}{ccccccccccc}
\hline \hline
Cr.P.& $\Omega_r$ & $\Omega_{\rm mg}$ & $\xi$ & $\Omega_m$ &  $\Omega_\sigma$ &
$w_{\rm eff}$ & $w_{\rm DE}$ &Existence& Stability & Eigenvalues\\
&&&&&&&&conditions&&\\
\hline
A&0 & 0  & 0  & $1-3/32\beta^2$ & $3/32\beta^2$ &  $-1/4$ & $-8\beta^2/3$ & $\beta^2\geq 3/32$ & Attractor & $-1$, $-3/4$, $-7/4$ \\

B& $1-3/32\beta^2$& 0  & 0  & 0 & $3/32\beta^2$ & 1/3 & 1/3 & $\beta^2\geq 3/32$ & Saddle & $-1$, 1,7/4 \\

C&0 & $1-3/32\beta^2$  & 0 & 0 & $3/32\beta^2$ & 0  & 0 & $\beta^2\geq 3/32$ &  Saddle  & $-1$, $-1$, 3/4 \\

 \hline \hline

\end{tabular}}
\end{center}
\end{table*}

Next by differentiating Eq.~(\ref{frd1}) we have
\begin{eqnarray}
  \frac{\dot{H}}{H^2} \Eqn{=} -\frac{1}{2(1-\Omega_\sigma-2\Omega_{\text{gal}})}\Bigg\{3\Omega_m\(1+\frac{F_1'}{12F_2}\)+4\Omega_r \nn \\ &&
 +\Omega_{\rm mg}\left[3+\xi\(\frac{F_1'}{F_1}-\frac{F_2'}{F_2}\)\right]+
 \xi(2\Omega_\sigma+3\Omega_{\rm gal}) \nn \\
&& \times \(\frac{F_1''}{F_1}-\frac{F2'}{F_2}\)
 -\frac{\lambda}{4\beta}y^2\frac{F_1'}{F_2}\Bigg\}\,,
 \label{hdot}
\end{eqnarray}
where
\begin{eqnarray}
 \Omega_m \Eqn{=} \frac{\rho_m}{3H^2\Mpl^2} \,, \\
 \Omega_r \Eqn{=} \frac{\rho_r}{3H^2\Mpl^2} \,,
\end{eqnarray}
are matter and radiation density parameters and
\begin{eqnarray}
 y \Eqn{=} \frac{\sqrt{V}}{\sqrt{3}H\Mpl}\,, \\
 \lambda \Eqn{=} -\Mpl \frac{{\rm d}V/{\rm d}\sigma}{V}\,.
\end{eqnarray}

It would be convenient to express the the effective
equation of state parameter $w_{\text{eff}}$ and the dark energy
equation of state parameter $w_{\text{DE}}$, respectively, as
\begin{eqnarray}
w_{\text{eff}} \Eqn{=} -1-\frac{2}{3}\frac{\dot H}{H^2}\,,
\label{Weff} \\
w_{\text{DE}} \Eqn{=}
\frac{w_{\rm eff}-\frac{1}{3}\Omega_r}{\Omega_{\rm DE}}\,,
\label{WDE}
\end{eqnarray}
where $\Omega_{\rm DE}$ is the dimensionless density parameter of
the dark energy components, i.e,
\begin{equation}
 \Omega_{\rm DE}=\Omega_\sigma+\Omega_{\rm mg}+\Omega_{\rm gal}+y^2 \,.
\end{equation}

In the following subsection, we shall discuss the cosmological
dynamics of the model under consideration using the above framework.

\subsubsection{Evolution for $c_3=0$ and $V(\sigma)=0$}
\label{case1}
Let us first consider the action (\ref{Eaction5}) without invoking
the potential and the galileon term. As we said earlier that though
the potential term is necessary to get the de Sitter solution for mass
varying nonlinear massive gravity, we are here interested in to study
the effect of the conformal coupling on the cosmological evolution. It
would be instructive to cast the evolution equations in the
autonomous system by defining the dimensionless variables $\xi$,
$\Omega_{\rm mg}$ and $\Omega_r$ as
\begin{eqnarray}
 \frac{\rm d\xi}{\rm d \ln a} \Eqn{=} -\xi \,,
 \label{eqxi1}\\
 \frac{\rm d\Omega_{\rm mg}}{\rm d \ln a} \Eqn{=} -\Omega_{\rm mg}\left[3+\xi\(\frac{F_1'}{F_1}-\frac{F_2'}{F_2}\)+2\frac{\dot{H}}{H^2}\right] \,,
 \label{eqLambda1}\\
 \frac{\rm d\Omega_r}{\rm d \ln a} \Eqn{=}
-2\Omega_r \(2+\frac{\dot H}{H^2}\) \,.
 \label{eqr1}
\end{eqnarray}
Looking at the above equations, it is seen that for every fixed
points we have $\xi=0$ such that at the fixed points,
$\Omega_\sigma=\frac{3}{32\beta^2}$ giving rise to $\beta^2\geq
3/32$. For estimates, we shall take the generic value of
$\beta \sim 1$.

It should be noted that for $\beta\to 0$, we find $\Omega_\sigma\to
\infty$, which means that we do not have viable cosmology. This is
obvious because in the limit $\beta\to 0$, the model under
consideration reduces to the $dRGT$, which does not support the
spatially flat FLRW cosmology \cite{D'Amico:2011jj}.
Let us note that in the Ref. \cite{Gumrukcuoglu:2013nza}, apart from the $\sigma$ field
which defines the variable mass, the Stuckelberg field $f(t)$ is
itself dynamical giving rise to a wider phase space than the case
under consideration.
Critical points and their nature of stability for the autonomous
system under consideration are written in Table~\ref{tab1}. The
stability analysis shows that massive gravity mimics matter during
the late time giving rise to an effective equation of state
parameter $\omega_{\rm eff}=-0.25$ which is an attractor of the
dynamics(see Table~\ref{tab1}). Critical point (A)
 in the table is the only attractor in this case. This stable point
corresponds to $\Omega_{\rm mg}=0$ and $\Omega_m\simeq 1$ ($\beta \sim
1$). The matter acquires here a slightly negative equation of state
due to its interaction with the other components in the model.

\begin{figure}[ht]
\begin{center}
\mbox{\epsfig{figure=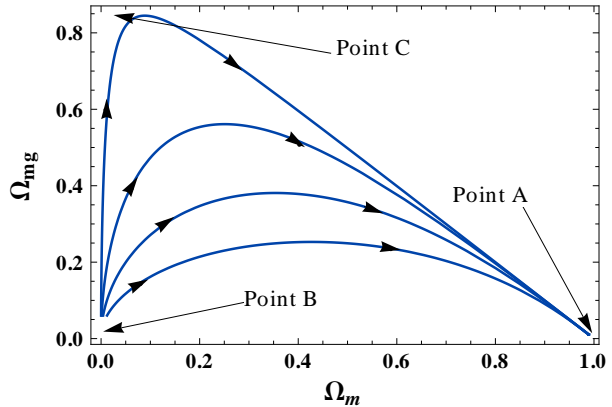,width=8.1cm,angle=0}}
\caption{Phase portrait for the dynamical system corresponding t
equations (\ref{eqxi1})-(\ref{eqr1}) is represented by trajectories
in the $\Omega_{\rm mg}$-$\Omega_m$ plane for the parameter choice
$\alpha_3=1$, $\alpha_4=1$ and $\beta=3$. Different trajectories are
for different initial conditions. All trajectories are seen to
converge to the point $A$ which represents matter dominated era as
an attractor solution for the case we have considered. The figure
also clearly shows that the massive gravity dominated era is an
intermediate era which is a saddle point.}
\label{Fig1}
\end{center}
\end{figure}

\begin{figure}[ht]
\begin{center}
\mbox{\epsfig{figure=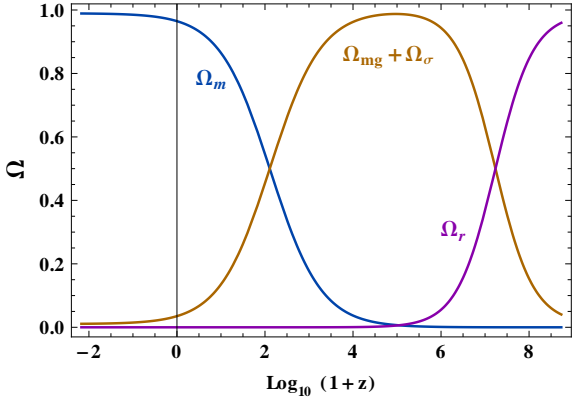,width=7.5cm,angle=0}}
\caption{This figure shows the evolution of the different components
of the energy density for the case $c_3=0$ and $V(\sigma)=0$ i.e,
for the evolution equations (\ref{eqxi1})-(\ref{eqr1}). Massive
gravity dominated era is an saddle point and an intermediate state.
During evolution, the system exits from the massive gravity era and
enters into the matter dominated phase which is an attractor. This
figure is plotted for the parameter choices $\alpha_3=1$,
$\alpha_4=1$ and $\beta=3$.}
\label{densitynp}
\end{center}
\end{figure}

\begin{figure}[ht]
\begin{center}
\mbox{\epsfig{figure=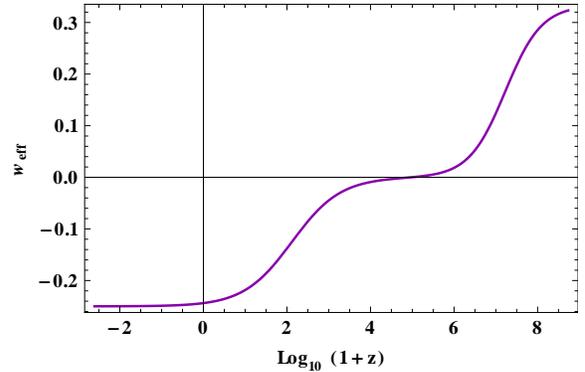,width=7.6cm,angle=0}}
\caption{This figure shows the evolution of the effective equation of state
($w_{\rm eff}$) for the case $c_3=0$ and $V(\sigma)=0$ i.e, for the
evolution equations (\ref{eqxi1})-(\ref{eqr1}). This figure clearly
shows that $w_{\rm eff}$ approaches $-0.25$ at late times which is
matter dominated era,  an attractor of the system.  Figure is
plotted for the parameter choices, $\alpha_3=1$, $\alpha_4=1$ and
$\beta=3$.}
\label{eosnp}
\end{center}
\end{figure}

Clearly we do not have dark energy solution in this case.
Fig.~\ref{Fig1} confirms the stability nature of the fixed point.
All trajectories are seen converging towards the point $A$.
Fig.~\ref{densitynp} shows the cosmological evolution of the
different components of the energy density. It also shows that the
massive gravity dominated era is an intermediate state and a saddle
point. Fig.~\ref{eosnp} also supports the results depicted in
Table~\ref{tab1} and clearly shows that during the late time, which
is matter dominated, the
effective equation of state ($w_{\rm eff}$) becomes $-0.25$ and remains in that
state since it is an attractor solution.
During the numerical analysis to keep $\Omega_\sigma$ small
we  took $\beta=3$ to plot this phase-portrait. Our analysis shows
that conformal coupling alone can not ensure a viable cosmology. We
might look for other possibilities to correct this situation.

\subsubsection{Evolution for $c_3\neq 0$ and $V(\sigma)=0$}
\label{case2}

\begin{table*}[ht]
\caption[crit]{The critical points of the autonomous system
(\ref{eqxi1})-(\ref{eqr1}) and (\ref{eqgalg}), their nature of stability, the
eigenvalues of the perturbation matrix are given here. The values of effective equation of
state ($w_{\text{eff}}$) and dark energy equation of state ($w_{\text{DE}}$) are also given.}
\label{tab2}
\begin{center}
\resizebox{\textwidth}{!}{%
\begin{tabular}{ccccccccccc}
\hline \hline
Cr.P.& $\Omega_r$ & $\Omega_{\rm mg}$ & $\xi$ & $\Omega_{\rm gal}$ & $\Omega_m$ &  $\Omega_\sigma$ &
$w_{\rm eff}$ & $w_{\rm DE}$ & Stability & Eigenvalues\\
\hline
D & $1-3/32\beta^2$& 0  & 0  & 0 & 0 & $3/32\beta^2$ & 1/3 & 1/3 & Saddle &$-1$,1,$7/4$,$-4$ \\

E &0 & 0  & 0  & 0 &  $1-3/32\beta^2$ & $3/32\beta^2$ &   $-1/4$ & $-8\beta^2/3$ & Attractor
&$-1$,$-9/4$,$-3/4$,$-7/4$ \\

F &0 & $1-3/32\beta^2$  & 0 & 0 & 0 & $3/32\beta^2$  & 0 & 0 &  Saddle  & $-1$,$-1$,$3/4$,$-3$ \\

G & 0 & 0 & 0 & $1-3/32\beta^2$ & 0 & $3/32\beta^2$  & $-1$ & $-1$ &  Attractor  & $-1$,$-3$,$-4$,$-9/4$ \\
 \hline \hline

\end{tabular}}
\end{center}
\end{table*}

We next examine the case for which $c_3\neq 0$ but without
invoking the potential term. We again need to set up the autonomous
system which can be done by introducing one more dimensionless
variable $\Omega_{\rm gal}$  to quantify the contribution of the
galileon term in addition to the variables considered in the
previous subsection. Consequently, we have the following evolution
equation for $\Omega_{\rm gal}$,

\begin{align}
\frac{\rm d\Omega_{gal}}{\rm d \ln a}&=-\Omega_{\text{gal}}\left[3\xi\(\frac{F_1''}{F_1}-\frac{F_2'}{F_2}\)-
 2\frac{\dot{H}}{H^2}\right]\,.
 \label{eqgalg}
\end{align}
Here Eq.~(\ref{eqgalg}) along with Eqs.~(\ref{eqxi1})--(\ref{eqr1}) will form the autonomous system.

In this case, we can define the total dark energy density
parameter as
\begin{equation}
 \Omega_{\text{DE}}=\Omega_\sigma+\Omega_{\text{gal}}+\Omega_{\rm mg}\,.
\end{equation}

\begin{figure}[ht]
\begin{center}
\mbox{\epsfig{figure=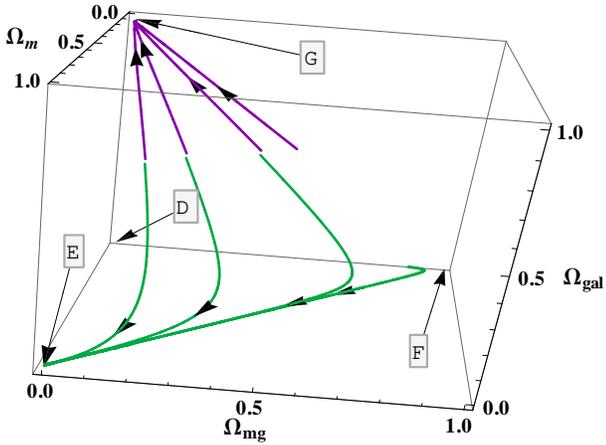,width=8cm,angle=0}}
\caption{Phase portrait of the dynamical system
(\ref{eqxi1})-(\ref{eqr1}) and (\ref{eqgalg}) for the parameter choices
$\alpha_3=1$,$\alpha_4=1$ and $\beta=3$. Brown lines represent the
trajectories which go towards the de Sitter solution without
following the thermal history of the universe. Green lines represent
the trajectories which go towards the attractor point $E$ through
the matter and radiation eras but can not reach the de Sitter
solution.} \label{Fig2}
\end{center}
\end{figure}

Table~\ref{tab2} displays the fixed points and their stability
analysis. It is clear from the table that in this case, there are
two attractors, one of them, $(E)$, represents the matter dominated
phase whereas the second attractor $(G)$ corresponds to de Sitter
solution. Keeping in mind the generic values of the coupling
$\beta\sim 1$, it follows from Table~\ref{tab2} that the de Sitter
is contributed by the galileon term ($\Omega_{\rm gal}\simeq 1$ and
$\Omega_m=0$) whereas on the other hand, in case of the attractor
(D), $\Omega_{\rm gal}=0$ and $\Omega_m\sim 1$. It should be emphasized
that both the attractors exist in the phase space simultaneously and
this becomes problematic. As a result, though we have a de Sitter
solution in this model but it can not be approached during
evolution. As universe exits the radiation era and enters the matter
dominated phase, it never leaves it as the latter is an attractor of
the dynamics. However, if we begin evolving the system very close to
de Sitter, we can have late time cosmic acceleration but it requires
unnatural fine tunings. Fig.~\ref{Fig2} is the phase portrait of the
dynamical system under consideration. Fig.~\ref{Fig2} confirms the
results of Table~\ref{tab2} and it clearly shows that since we have
two attractor solutions, the system will choose one of them
depending upon the initial conditions. It is also clear from the
figure that if we pick up initial conditions for which the system
chooses the attractor solution at point $G$ then we can not have
matter and radiation dominated eras and the system remains in the de
Sitter phase forever. To have the matter and radiation eras we need
to choose initial conditions appropriately so that the system
chooses the attractor ($E$) which is naturally reached during
evolution. But in that case we cannot reach the de Sitter solution
which is screened out here by the matter phase.

It is clear that formally we can realize the de Sitter solution at the cost of
minimal completion of scalar field by lower order galileon term
$L_3$. We did not invoke $L_4$ which by itself can give rise to the de
Sitter solution in the standard framework \cite{Gannouji:2010au,Ali:2010gr}.

\subsubsection{Evolution for $c_3=0$ and $V(\sigma)\neq 0$}
\label{case3}

\begin{table*}[ht]
\caption[crit]{The critical points of the autonomous system (\ref{eqxi1})-(\ref{eqr1}) and (\ref{eqyp}),
their nature of stability, the
eigenvalues of the perturbation matrix are given here. The values of effective equation of
state ($w_{\text{eff}}$) and dark energy equation of state ($w_{\text{DE}}$) are also given.}
\label{tab3}
\begin{center}
\resizebox{\textwidth}{!}{%
\begin{tabular}{ccccccccccc}
\hline \hline
Cr.P.& $\Omega_r$ & $\Omega_{\rm mg}$ & $\xi$ & y & $\Omega_m$ &  $\Omega_\sigma$ &
$w_{\rm eff}$ & $w_{\rm DE}$ & Stability & Eigenvalues\\
\hline
H & $1-\frac{3}{32\beta^2}$& 0  & 0  & 0 & 0 & $\frac{3}{32\beta^2}$ & 1/3 & 1/3 & Saddle
& -1,1,$\frac{7}{4}$,$2-\frac{3\lambda}{8\beta}$\\

I &0 & 0  & 0  & 0 &  $1-\frac{3}{32\beta^2}$ & $\frac{3}{32\beta^2}$ &   $-1/4$ & $-8\beta^2/3$ & Saddle for
&-1,$-\frac{3}{4}$,$-\frac{7}{4},\frac{3 (3\beta -\lambda)}{8 \beta}$ \\
 &&&&&&&& & $\lambda<3\beta$ &\\

J &0 & $1-\frac{3}{32\beta^2}$  & 0 & 0 & 0 & $\frac{3}{32\beta^2}$  & 0 & 0 & Saddle
& -1,-1,$\frac{3}{4}$,$\frac{3}{8} \left(4-\frac{\lambda }{\beta }\right)$\\

K & 0 & 0 & 0 & $\pm \left(1-\frac{3}{32\beta^2}\right)^\frac{1}{2}$ & 0 & $\frac{3}{32\beta^2}$  & $-1+\frac{\lambda}{4\beta}$ &
$-1+\frac{\lambda}{4\beta}$ & Attractor for   &
-1,$-4+\frac{3\lambda }{4\beta }$,$-3+\frac{3\lambda}{4\beta}$, \\
&&&&&&&&& $\lambda<3\beta~~\rm{for} \beta>0$ &$-\frac{9}{4}+\frac{3\lambda}{4\beta}$ \\
&&&&&&&&& $\lambda<16\beta/3~~\rm{for} \beta<0$ & \\
 \hline \hline

\end{tabular}}
\end{center}
\end{table*}

Now let us explore the case with $c_3=0$ and non zero potential. This
system is quite similar to the mass varying massive gravity model
except that here we have a non minimal coupling between the matter
field and the $\sigma$ field in the Einstein frame (minimal coupling
case is studied in Refs.~\cite{Saridakis:2012jy,Leon:2013qh,Cai:2012ag,Hinterbichler:2013dv}).
In the previous section we have observed that the non minimal coupling
modifies the matter phase and does not naturally gives rise to late
time acceleration, it will be interesting here to compare the
minimal and non minimal cases.

To check for cosmological evolution, let us consider an exponential
potential $V(\sigma)=V_0 \mathrm{e}^{-\lambda\sigma/\Mpl}$. Also consider the
dimensionless variable $y$ along with the dimensionless variables
$\xi$, $\Omega_{\rm mg}$ and $\Omega_r$ to form the autonomous
system. The evolution equation for $y$ has the form,
\begin{align}
\frac{\rm dy}{\rm d \ln a}&=y\left(\frac{\lambda}{8\beta}\frac{F_1'}{F_2}-\frac{\dot{H}}{H^2}\right)\,,
 \label{eqyp}
\end{align}
where $\dot{H}/H^2$ is given by Eq.~(\ref{hdot}) considering $c_3=0$.
For this case the total energy density of the
dark energy is given by
\begin{equation}
 \Omega_{\rm DE}=\Omega_\sigma+\Omega_{\rm mg}+y^2 \,.
\end{equation}

The Eq.~(\ref{eqyp}) with Eqs.~(\ref{eqxi1})--(\ref{eqr1}) forms
the autonomous system. Fixed points and their nature of stability
are indicated in Table~\ref{tab3}. In this case we have only one
attractor which is a de Sitter solution. We should note that like
other cases the matter phase is disturbed here also due to the
effective pressure of the matter. Now to have small values of
$\Omega_\sigma$ we have to set the value of $\beta$ at least such
that $\beta \sim 1$. This tells us that $w_{\rm eff}\backsimeq -1$
implies that $\lambda$ takes small numerical values.

\begin{figure}[ht]
\begin{center}
\mbox{\epsfig{figure=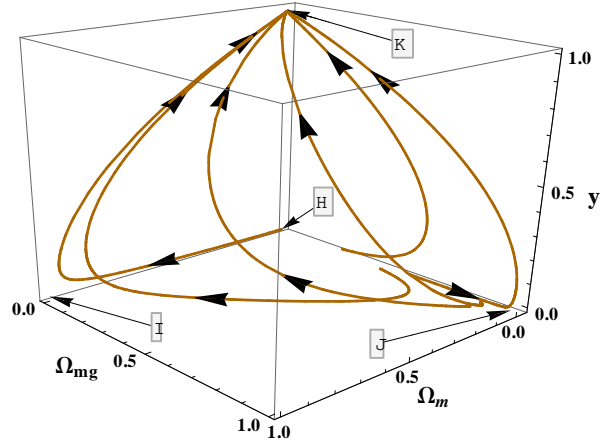,width=8cm,angle=0}}
\caption{Phase portrait of the dynamical system (\ref{eqxi1})-(\ref{eqr1}) and
(\ref{eqyp}) for the parameter choices $\alpha_3=1$,$\alpha_4=1$,
$\beta=3$ and $\lambda=0.001$. Attractor solution is represented by
the point $K$ which is a de Sitter solution and represents the
potential dominated era. It is clearly seen that the massive gravity
dominated era is an intermediate state which is represented by the
saddle point $J$. An important thing to notice is that the matter
phase, which is represented by the point $I$, unlike the other two
cases becomes a saddle point.}
\label{Fig3}
\end{center}
\end{figure}

Fig.~\ref{Fig3} shows the stability of critical points of the
autonomous system (\ref{eqxi1})--(\ref{eqr1}) and (\ref{eqyp}) which
is consistent with the results described in the Table~\ref{tab3}. In
Fig.~\ref{Fig3} the points $H$, $I$, $J$ and $K$ are represented by
the points $(\Omega_m,\Omega_{\rm mg}, y)\equiv (0,0,0)$, $(\sim
1,0,0)$, $(0,\sim 1,0)$ and $(0,0,\sim 1)$ respectively. We can see
that all trajectories are going towards the attractor solution i.e,
point $K$. Fig.~\ref{Fig3} and Table~\ref{tab3} show that the
stability nature of this system is similar to the mass-varying
massive gravity model without any conformal coupling.

\begin{figure}[ht]
\begin{center}
\mbox{\epsfig{figure=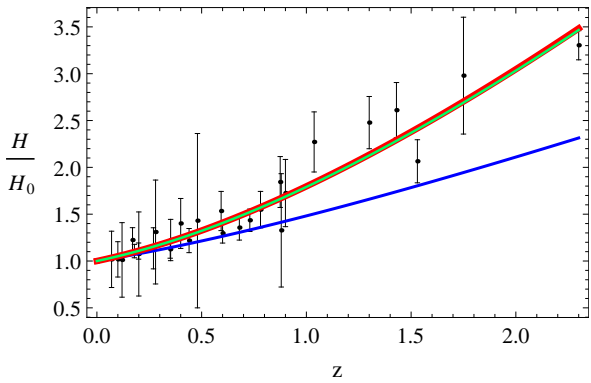,width=7.8cm,angle=0}}
\caption{Figure
shows the evolution of the normalized Hubble parameter.
Observational data points for $H/H_0$ calculated from the available
data of $H(z)$ are shown as the black dots with their $1\sigma$
error bars. Blue, red and green lines  represent the evolution of
$H/H_0$ in non minimally coupled mass-varying massive gravity,
minimally coupled mass-varying massive gravity and $\Lambda$CDM
models respectively. To show the overlapping between the red and the
green line we have made the red line thicker than the green line. To
plot the blue and the green line we have chosen
$\alpha_3=1$,$\alpha_4=1$, $\beta=3$ and $\lambda=0.001$.} \label{Fig4}
\end{center}
\end{figure}

Now let us compare these results with the minimal coupling case. For
the minimal coupling case the fixed points and their nature of
stability remains same but what changes is the $w_{\rm eff}$ for the
matter dominated era and it changes to standard equation of state
for matter. Also the $w_{\rm DE}$ approaches $0$ during matter
dominated era. Contribution from
massive gravity for all the cases behaves like matter giving rise to
conventional equation of state for matter. So the coupling between
matter and the field $\sigma$ actually
removes the degeneracy between  the two fixed points representing
the matter dominated era and massive gravity dominated era.

As for the effect of the coupling on the evolution let us check the
evolution of Hubble parameter. In Fig.~\ref{Fig4} we have compared
the evolution of the Hubble parameter for non minimal coupling case
with the minimal coupling case and $\Lambda$CDM model. We have used
the available data for Hubble parameter $H(z)$
\cite{Simon:2004tf,Stern:2009ep,Moresco:2012by,Zhang:2012mp,Blake12,Busca12,Chuang2012b}
and their $1\sigma$ error bars. {}From the data of $H(z)$, we have
calculated normalized Hubble parameter, i.e, $H/H_0$ where $H_0$ is
the present value of the Hubble parameter. We have taken the value
of $H_0$ from PLANCK 2013 results \cite{Ade:2013zuv} and we have set
$H_0=67.8~~\rm Km/Sec/Mpc$. We have also estimated the error in
$H/H_0$ from the given data. During the evolution we have set the
present value of $\Omega_m$ to $0.315$ which is taken from the
PLANCK 2013 results \cite{Ade:2013zuv}. Fig.~\ref{Fig4} shows that
the evolution of the normalized Hubble parameter in $\Lambda$CDM and
minimally coupled case are very similar and also consistent with the
available data. But the evolution of the Hubble parameter in the non
minimally coupled case defers a lot from the $\Lambda$CDM and
minimally coupled case and is  inconsistent with the present data.
The deviations of the Hubble parameter changes the evolution of the
matter, radiation and dark energy density parameters.

In the three cases discussed above, the limit $\beta\to 0$ is
inconsistent which is obvious because in this limit we recover the
original $dRGT$ nonlinear massive gravity which cannot give rise to the
flat FLRW solution. The model under consideration can give rise to the
flat FLRW cosmology with non minimal coupling only. To make the
theory consistent with the $\beta\to 0$ limit we may consider a
potential term of the $\sigma$ field in the massive gravity part.

\section{Non-minimal coupling in quasi-dilaton nonlinear massive gravity}
\label{quasidilaton}

In the models discussed in the preceding section, one needs a
potential term to have a de Sitter solution irrespective of the
conformal coupling term. Though the addition of the cubic galileon
term gives a de Sitter solution without potential it is screened by
the massive gravity dominated era which is also an attractor
solution. So in any case massive gravity term in the varying mass
nonlinear massive gravity does not give rise to a viable de Sitter
solution. But in quasi-dilaton nonlinear massive gravity we get self
acceleration from the massive gravity part
\cite{D'Amico:2012zv,Gannouji:2013rwa} without adding any potential
term or a galileon. Now let us consider non-minimal coupling in
quasi-dilaton nonlinear massive gravity \cite{D'Amico:2012zv} and
consider the following action in the Einstein frame
\begin{align}
 \label{Eaction6}
 \mathcal{S}_E &=\int
\rm d^4x\sqrt{-g}\bigg[\frac{\Mpl^2}{2}R-\frac{\omega}{2}g^{\mu \nu}
\partial_\mu \sigma \partial_\nu
\sigma \nn \\ & -\frac{m^2\Mpl^2}{8}\left(4\,\mathcal{U}_2(\bar{\Bbbk})+\alpha_3\mathcal{U}_3(\bar{\Bbbk})+
\alpha_4\mathcal{U}_4(\bar{\Bbbk})\right)\bigg]\nn \\
&+\int d^4x\sqrt{-g}\mathcal{L}_m(A^2g_{\mu \nu},\psi)\,,
 \end{align}
where
\begin{equation}
\bar{\Bbbk}^\mu_\nu=\delta^\mu_\nu-e^{(\beta_q+\beta)\sigma/\Mpl}
\sqrt{g^{\mu\alpha}\partial_\alpha\phi^a\partial_\nu\phi^b\eta_{ab}}\,,
\end{equation}
with $\beta_q=1$ for the the quasi-dilaton nonlinear massive
gravity described in \cite{D'Amico:2012zv}. Though for minimally
coupled case ($\beta=0$), we can absorb $\beta$ within the $\sigma$
field by redefining the field leaving the model unchanged.

The action (\ref{Eaction6}) can be motivated from the Jordan frame
action
\begin{eqnarray}
\label{Eaction7}
\mathcal{S}_J &=& \frac{\Mpl^2}{2} \int \rm d^4 x \sqrt{-\tilde{g}} \nn \\ &&
\bigg[ \mathrm{e}^{2\beta\sigma/\Mpl}
\(\tilde{R} +\frac{(6\beta^2-\omega)}{\Mpl^2}\tilde{g}^{\mu\nu}\partial_\mu\sigma\partial_\nu\sigma\) \nn \\
&&- \frac{m^2}{4}\mathrm{e}^{4\beta\sigma/\Mpl} \( 4\,\mathcal{U}_2(\Bbbk) +\alpha_3 \mathcal{U}_3(\Bbbk)+
 \alpha_4 \mathcal{U}_4(\Bbbk) \) \bigg] \nn \\ &&
+\sqrt{-\tilde{g}}\mathcal{L}_m(\tilde g_{\mu\nu}, \psi) \,,
\end{eqnarray}
where
\begin{equation}
\Bbbk^\mu_\nu=\delta^\mu_\nu-e^{\beta_q\sigma/\Mpl}
\sqrt{g^{\mu\alpha}\partial_\alpha\phi^a\partial_\nu\phi^b\eta_{ab}} \,.
\end{equation}

\subsection{Lagrangian in the decoupling limit}
\label{dlimit_qd}
The action (\ref{Eaction6}) is same as the quasi-dilaton nonlinear
massive gravity except we have a coupling between the $\sigma$ field
with the matter field which will modify the continuity equation of
matter. But even in the presence of the conformal coupling the
decoupling limit (\ref{dcplng}) Lagrangian is same as the original
model \cite{D'Amico:2012zv}. As for the local physics, we can have a
contribution from the conformal coupling  which  comes from the
matter conservation equation. But similar to the varying mass
nonlinear massive gravity,  the contribution from the conformal
factor contains a factor like $\beta/\Mpl$ which  makes the
contribution vanishing in the decoupling limit. Thus the
action (\ref{Eaction6}) preserves the local physics through the
Vainshtein mechanism.

\subsection{Cosmology}
\label{cosmology_qd}
By redefining the $\sigma$ field as $\sigma\to \sigma/(\beta_q+\beta)$ in
the FLRW background (\ref{metric}) the action (\ref{Eaction6}) becomes
\begin{eqnarray}
&& \hspace{-5mm}
\mathcal{S}_E = 3M_{Pl}^2\int {\rm d}t \biggl[-\frac{a\dot a^2}{N}+\frac{\omega_0}{6\Mpl^2}~ \frac{a^3\dot
\sigma^2}{N} \nn \\
&& \hspace{1mm}
+ m^2a^3\left(N G_1(\zeta)-\dot f a G_2(\zeta)\right)\biggr]+\mathcal{S}_m+\mathcal{S}_r \,,
\label{Eaction8}
\end{eqnarray}
where for the St\"{u}ckelberg scalars we consider the ansatz
\begin{align}
\phi^0=f(t)\,, ~~~ \phi^i=x^i \,,
\end{align}
and we have defined
\begin{align}
\label{G1def}
G_1(\zeta) &= (1-\zeta)\biggl[2-\zeta+\frac{\alpha_3}{4} (1-\zeta)(4-\zeta)
\nonumber \\
&+\alpha_4 (1-\zeta)^2\biggr]\,,\\
\label{G2def}
G_2(\zeta) &= \zeta (1-\zeta)\biggl[1+\frac{3}{4}\alpha_3 (1-\zeta)+\alpha_4
(1-\zeta)^2\biggr]\,, \\
\omega_0 &=\frac{\omega}{(\beta_q+\beta)^2}\,,
\end{align}
where $\zeta$ is defined as
\begin{equation}
 \zeta=\frac{\mathrm{e}^{\sigma/\Mpl}}{a}.
\end{equation}
In action (\ref{Eaction8}) $\mathcal{S}_m$ and $\mathcal{S}_r$
represent the matter and radiation actions respectively.
We should keep in mind that in the Einstein frame matter
is coupled with the $\sigma$ field and this will modify
the conservation equation of the matter.

Variation the action (\ref{Eaction6}) with respect to $f$
gives the constraint equation (as an essential feature of
any model of non-linear massive gravity)
\begin{align}
\label{eq:Constraint}
G_2(\zeta)=\frac{C}{a^4}\,,
\end{align}
where $C$ is a constant of integration.

Varying the action (\ref{Eaction6}) with respect to $N(t)$ and setting
$N(t)=1$ at the end
we get the Friedmann equation \cite{Gannouji:2013rwa}
\begin{align}
\label{eq:Fried}
3\Mpl^2H^2=\frac{\rho_m+\rho_r-3 m^2 \Mpl^2G_1}{1- \frac{\omega_0}{6}
\Bigl(1-4\frac{G_2}{\zeta G_2'}\Bigr)^2} \,,
\end{align}
where $\rho_m$ and $\rho_r$ are the matter and radiation energy density respectively.
The primes in $G$'s denote derivatives with respect to their argument
$\zeta$.

\begin{table*}[ht]
\caption[crit]{\label{qd} The real and physically meaningful
critical points of the autonomous system (\ref{eqr})-(\ref{eqxi}), their
existence conditions, the eigenvalues of the perturbation matrix, and the
deduced stability conditions. We also present the corresponding
values of the various density parameters, and the values of the
observables: deceleration parameter $q$, total equation-of-state
parameter $w_{\rm eff}$ and dark energy equation-of-state
parameter $w_{\rm DE}$.}
\begin{center}
\resizebox{\textwidth}{!}{%
\begin{tabular}{cccccccccccc}
\hline \hline Cr.P.&
$\Omega_r$ & $\Omega_\Lambda$ & $\zeta$ & $\Omega_m$ &  $\Omega_\sigma$ &$q$&
$w_{\rm eff}$ & $w_{DE}$ &Existence& Stability & Eigenvalues\\
&&&&&&&&&conditions&& \\
\hline
$P_1$ &$1-3\omega_0/2$ & 0  &0  & 0 & $3\omega_0/2$ &1& $1/3$ & $1/3$
&$0\leq\omega_0\leq2/3$& Saddle point & $-4$, 4, $1-3\beta_0$ \\
&&&&&&&&&& for all $\beta_0$ & \\

$P_2$ &0 & 0  & 0  & $1-3\omega_0/2$ & $3\omega_0/2$ &$(1+3\beta_0)/2$   &  $\beta_0$ & $2\beta_0/3\omega_0$
&$0\leq\omega_0\leq2/3$& Saddle point & $-4$, $3+3\beta_0$, \\
&&&&&&&&&& for $\beta_0>-1$ &  $-1+3\beta_0$  \\

$P_3$ &0 & $1-3\omega_0/2$  & 0 & 0 & $3\omega_0/2$ &-1   &  -1 & -1
&$0\leq\omega_0\leq2/3$&  Attractor  & $-4$, $-4$,  \\
&&&&&&&&&& for $\beta_0>-1$ & $-3(1+\beta_0)$\\

$P_{4\pm}$&$1-\omega_0/6$ & 0  & $\zeta_\pm$ & 0 & $\omega_0/6$ &1& $1/3$ & 1/3
&$0\leq\omega_0\leq6$,& Saddle point & $-4$,4,$1+\beta_0$\\
&&&&&&&&&  $0\leq\zeta_\pm$& for all $\beta_0$ & \\

$P_{5\pm}$&0 & 0  & $\zeta_\pm$ & $1-\omega_0/6$ & $\omega_0/6$ &$(1-\beta_0)/2$& $-\beta_0/3$ & $-2\beta_0/\omega_0$ &
$0\leq\omega_0\leq6$,  & Saddle point & $-4$,$-1-\beta_0$, \\
&&&&&&&&&  $0\leq\zeta_\pm$ & for $\beta_0<3$ & $3-\beta_0$ \\

$P_{6\pm}$&0 & $1-\omega_0/6$  & $\zeta_\pm$ & 0 & $\omega_0/6$  &-1 &-1 & -1
&$0\leq\omega_0\leq6$,   & Attractor  & $-4$,$-4$,$-3+\beta_0$ \\
&&&&&&&&& $0\leq\zeta_\pm$ & for $\beta_0<3$ & \\

$P_7$ &0 & 0  &1  & $1-\omega_0/6$ & $\omega_0/6$ &$(1-\beta_0)/2$   &  $-\beta_0/3$ & $-2\beta_0/\omega_0$
&$0\leq\omega_0\leq6$ & Attractor &  $-4$, $-1-\beta_0$, \\
&&&&&&&&&& for $\beta_0>-1$ & $-1-\beta_0$ \\

$P_8$ &$\Omega_r$ & $1-\omega_0/6$  & $1$ & 0& $\omega_0/6$  &$1$&$1/3$
&  $1/3$ &$0\leq\omega_0\leq6$&  Saddle  & $-4$,$1+\beta_0$,0 \\
&&-$\Omega_r$&&&&&&&& for $\beta_0>-1$ &
 \\
 \hline\hline
\end{tabular}}
\end{center}
\end{table*}

Similarly, variation with respect to $\sigma$ gives the dilaton
evolution equation
 \begin{equation}
 \frac{\omega_0}{a^3}\frac{d}{dt}(a^3\dot{\sigma})-3\Mpl
m^2\zeta\left(G_1'(\zeta)-a\dot{f}G_2'(\zeta)\right)
= -\frac{\beta_0}{\Mpl}\rho_m \,,
\end{equation}
where
\begin{equation}
 \beta_0=\frac{\beta}{\beta_q+\beta} \,.
 \label{beta0}
\end{equation}
Continuity equations for matter and radiation are same as
Eqs.~(\ref{contmat}) and (\ref{contrad}), respectively. Now using
Eq.~(\ref{eq:Constraint}) we can get an effective pressure for
the matter phase which is given by
\begin{equation}
 p_{m(\rm eff)}=-\frac{\beta_0}{3}\(1-\frac{4G_2}{\zeta G_2'}\)\,.
 \label{eff_pres_qd}
\end{equation}
Here we can also have an effective pressure for the matter phase
which is the contribution from the conformal coupling. Now let us
study how this effective pressure affects the cosmology of the
model. And to study this we need to form an autonomous system. Let
us first introduce the density parameters
\begin{align}
\Omega_m &=\frac{\rho_m}{3 \Mpl^2 H^2}\,, \\
\Omega_r &=\frac{\rho_r}{3 \Mpl^2 H^2}\,, \\
\label{OmegaLambda}
\Omega_{\Lambda} &=-\frac{m^2}{H^2}G_1 \,, \\
\Omega_\sigma &=\frac{\omega_0}{6} \left(1-4 \frac{G_2}{\zeta G_2'}
\right)^2 \,,
\end{align}
with which we can rewrite the Friedmann equation (\ref{eq:Fried}) as
 \begin{align}
\label{Fr1b}
\Omega_m+\Omega_r+ \Omega_\Lambda+\Omega_\sigma=1 \,.
\end{align}

We can now transform the above cosmological system into its
autonomous form, using only the dimensionless variables $\Omega_r$,
$\Omega_\Lambda$ and $\xi$, while (\ref{Fr1b}) will be used to
eliminate $\Omega_m$. Doing so we obtain
\begin{align}
\label{eqr}
\frac{{\rm d}\Omega_r}{{\rm d}\ln a} &= -2\Omega_r \left(2+\frac{\dot
H}{H^2}\right)\,, \\
\label{eqLambda}
\frac{{\rm d}\Omega_\Lambda}{{\rm d}\ln a} &=-2\Omega_\Lambda
\left( 2\frac{G_2 G_1'}{G_1 G_2'}+\frac{\dot H}{H^2}\right)\,, \\
\label{eqxi}
\frac{{\rm d}\xi}{{\rm d}\ln a} &=-4\frac{G_2}{G_2'}\,,
\end{align}
where the combination $\frac{\dot H}{H^2}$ can be acquired differentiating
the Friedmann equation (\ref{eq:Fried}) as
\begin{eqnarray}
&& \hspace{-10mm}
\frac{\dot H}{H^2} = \frac{1}{6-\omega_0\(1-4\frac{G_2}{\zeta G_2'}\)^2}  \nn \\ && \times
\Biggl\{-9\Omega_m\left[1-\frac{\beta_0}{3}\(1-\frac{4G_2}{\zeta G_2'}\)\right]- 12\Omega_r \nn \\ &&
-12\frac{G_2}{G_2'}\left[\frac{G_1'}{G_1
}\Omega_\Lambda+\frac{\omega_0}{6}\frac{\rm d}{{\rm d}\zeta}\(1-4\frac{G_2}{\zeta
G_2'}\)^2\right]\Biggr\}\,.
\label{HdotH2}
\end{eqnarray}

\begin{figure}[ht]
\begin{center}
\mbox{\epsfig{figure=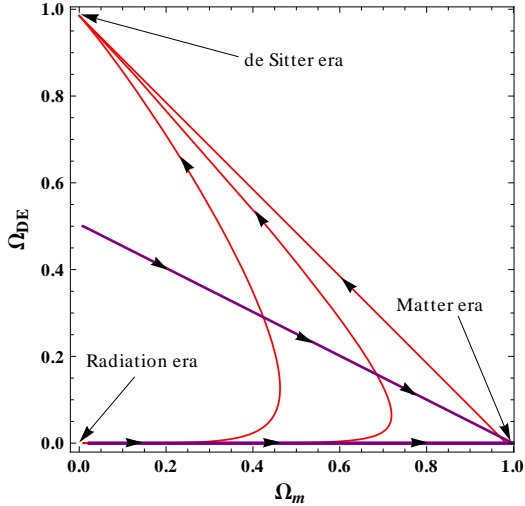,width=7cm,angle=0}}
\caption{Phase
portrait for the autonomous system (\ref{eqr})-(\ref{eqxi}).
Different  trajectories correspond to different initial conditions.
Red lines correspond to the initial conditions for which we get late
time acceleration $e.g$, points correspond to $\zeta=0$ or
$\zeta=\zeta_\pm$. Purple lines correspond to the trajectories for
which $\zeta \approx 1$ and gives matter dominated phase as the late
time solution. To plot the trajectories, we have chosen
$\alpha_3=1$, $\alpha_4=1$, $\omega_0=0.01$, $\beta_0=0.01$.}
\label{pp_qd}
\end{center}
\end{figure}

\begin{figure}[ht]
\begin{center}
\mbox{\epsfig{figure=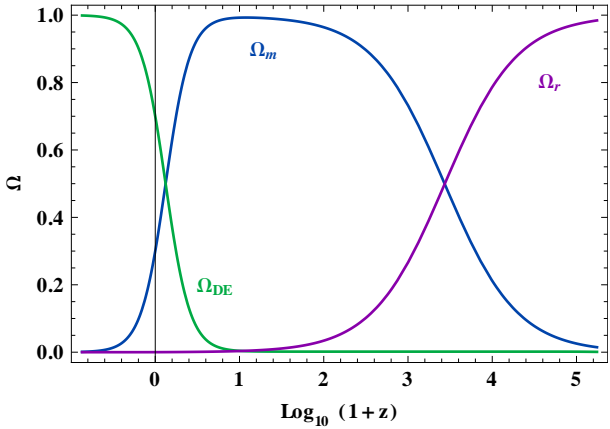,width=8.05cm,angle=0}}
\caption{Cosmological evolution of different density parameters are shown here for the parameter choices
$\alpha_3=1$, $\alpha_4=-4$, $\omega_0=0.01$, $\beta_0=0.01$. This figure shows that late time is
dark energy dominated for the cosmological evolution of the autonomous system (\ref{eqr})-(\ref{eqxi}).}
\label{density_qd}
\end{center}
\end{figure}

\begin{figure}[ht]
\begin{center}
\mbox{\epsfig{figure=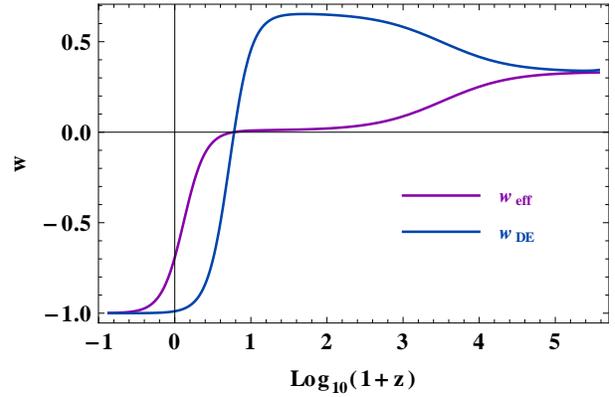,width=8.05cm,angle=0}}
\caption{Cosmological evolution of $w_{\rm eff}$ and $w_{\rm DE}$
are shown here for the parameter choices $\alpha_3=1$,
$\alpha_4=-4$, $\omega_0=0.01$, $\beta_0=0.01$. This figure shows
that the model under consideration gives late time acceleration as
an attractor solution. Figure correspond evolution described by the
autonomous system (\ref{eqr})-(\ref{eqxi}).}
\label{eos_qd}
\end{center}
\end{figure}

The effective equation of state parameter $w_{\rm eff}$, the deceleration parameter
$q$ and the dark energy equation of state parameter $w_{\rm DE}$
are defined as
\begin{eqnarray}
w_{\rm eff} \Eqn{=} -1-\frac{2}{3}\frac{\dot
H}{H^2}\,,
\label{Weff} \\
\label{qcase1}
q \Eqn{=} -1-\frac{\dot
H}{H^2}=\frac{1}{2}+\frac{3}{2}w_{\rm eff}\,, \\
w_{\rm DE} \Eqn{=} \frac{w_{\rm eff}-w_r\Omega_r}{\Omega_\Lambda+
\Omega_\sigma}\,,
\label{WDE}
\end{eqnarray}
where the combination $\dot {H}/H^2$ is given by Eq.~(\ref{HdotH2}) and
$w_r=1/3$.

The critical points of the above autonomous system are extracted setting
the left hand sides of equations (\ref{eqr})--(\ref{eqxi}) to zero. In
particular, equation (\ref{eqxi}) implies that at the critical points
$G_2(\zeta)=0$, which according
to (\ref{G2def}) gives $\zeta=0$, $\zeta=1$ and
\begin{align}
\label{xipm}
\zeta_\pm &=1+\frac{3\alpha_3\pm \sqrt{9\alpha_3^2-64\alpha_4}}{8\alpha_4}
\,.
\end{align}

To be physical, the critical points have to possess
$0\leq\Omega_r\leq1$, $0\leq\Omega_\Lambda+\Omega_\sigma\leq1$,
$0\leq\zeta$ and of course $\zeta\in {\cal{R}}$. Now, in order for
$\zeta_\pm\in {\cal{R}}$, we require $9\alpha_3^2-64\alpha_4\geq0$
while $0\leq\zeta$ leads to the necessary conditions for $\alpha$'s
using Eq.~(\ref{xipm}). All the fixed points and their nature of
stability along with the values of effective equation of
state ($w_{\rm eff}$) and dark energy equation of state ($w_{\rm DE}$)
are given in Table~\ref{qd}. Fig.~\ref{pp_qd} shows the trajectories
on the $\Omega_m$--$\Omega_{\rm DE}$ plane. For $\zeta=1$ the matter
phase is an attractor point which is represented in the
Fig.~\ref{pp_qd} by the purple lines. But for $\zeta=0$ or
$\zeta=\zeta_\pm$ we have late time acceleration as the attractor
solution and it is represented in the Fig.~\ref{pp_qd} by the red
lines.

To have viable thermal history of the universe we need to have
$\Omega_{\rm DE}\leq 0.01$ during the radiation dominated era which
gives a constraint on the value of $\omega_0$ ($\omega_0\leq 0.02/3$)
for $\zeta=0$ and $\omega_0\leq 0.06$ for the point
$\zeta=\zeta_\pm$. Though this is a rough estimation on the
constraint on $\omega_0$. We can have full constraint on $\omega_0$
by using the cosmological data sets. We also see from
Table~\ref{qd} that the conditions for stability of the fixed points and
the values of the dimensionless variables do not depend on
$\alpha$'s. But to have $\Omega_\Lambda>0$ for the points $P_3$ and
$P_{6\pm}$ we can have $2+\alpha_3+\alpha_4<0$ for the point $P_3$,
$\alpha_3>0$ and $0<\alpha_4<\alpha_3^2/8$ for the point $P_{6+}$
and $\alpha_3<0$ and $0<\alpha_4<\alpha_3^2/8$ for the point
$P_{6-}$. These conditions are needed to have viable evolution of
the cosmological parameters. But since the stability does not imply
explicit dependence on $\alpha$'s, we can not have constraints on
$\alpha$'s from the the observational data sets \cite{Gannouji:2013rwa}
because maintaining the conditions on
$\alpha$'s if we change their values, cosmology does not change
appreciably. Fig.~\ref{density_qd} depicts that the model under
consideration gives proper cosmological sequences starting from the
radiation dominated era to the late time cosmic acceleration
maintaining the proper matter dominated era. Cosmological evolution
of the effective equation of state ($w_{\rm eff}$) and dark energy
equation of state ($w_{\rm DE}$) are shown in Fig.~\ref{eos_qd}
which clearly shows that the present era is dark energy dominated.

It is clear from Table~\ref{qd} that due to the presence of
conformal coupling, we have a different evolution of the dark energy
equation of state ($w_{\rm DE}$) than the one  in the minimally
coupled quasi-dilaton nonlinear massive gravity
\cite{DeFelice:2012mx, Gannouji:2013rwa}. For the case under
consideration, $w_{\rm DE}$ has a non zero value for $\beta_0\neq 0$
during matter dominated era (see Fig.~\ref{eos_qd}). This figure is
plotted such that we can reach the attractor point $P_3$ during the
late time. Table~\ref{qd} also tells that for point $P_3$,
$\beta_0/\omega_0\sim 1$, during  matter era ($w_{\rm DE}\sim 2/3$)
which is confirmed numerically (see Fig.~\ref{eos_qd}). In
minimally coupled case, the value of $w_{\rm DE}$ during the matter
era is zero \cite{Gannouji:2013rwa} which can be checked from the
Table~\ref{qd} by putting $\beta_0=0$. Hence looking at the
evolution of $w_{\rm DE}$, we can differentiate between the
minimally coupled and non-minimally coupled quasi-dilaton nonlinear
massive gravity. It is really interesting to note (see Table~
\ref{qd}) that in case of the attractor represented by $P_{6\pm}$,
the parameter $w_{\rm DE}$ can reach super negative values before
reaching de Sitter during evolution depending upon the values of
$\beta_0$ and $\omega_0$. We also emphasize that the current data
allows phantom acceleration which is realized in our model as a
transient phenomenon$-$ {\it a phantom phase without a phantom
field}. The non-minimal coupling is responsible for the transient
phantom behavior, see Fig.~\ref{phantom}.

\begin{figure}[ht]
\begin{center}
\mbox{\epsfig{figure=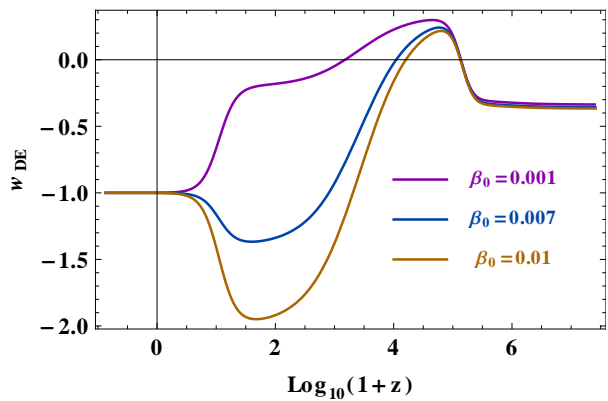,width=8.05cm,angle=0}}
\caption{Cosmological evolution of $w_{\rm DE}$ for the point
$P_{6+}$are shown here for the parameter choices $\alpha_3=1$,
$\alpha_4=0.115$, $\omega_0=0.01$ and for $\beta=0.01$, $0.007$ and
$0.001$. This figure shows that phantom phase  can be achieved
during the intermediate state of cosmological evolution if the
universe chooses such initial conditions for which it reaches the
point $P_{6\pm}$. This picture is drawn by evolving the autonomous
system (\ref{eqr})-(\ref{eqxi}).}
\label{phantom}
\end{center}
\end{figure}

\subsection{Observational Constraints}
\label{data_analysis}
We now get to the investigations on constraints we can have on the
model parameters from the observational data sets. The model under
consideration has the parameters $\alpha_3$, $\alpha_4$, $\beta_0$,
$\omega_0$, $m$, $\Omega_{m0}$, $C$ and $H_0$. To obtain the
observational constraints on the model parameters we have used the
observational data sets of Type Ia Supernovae (SNIa)
\cite{Suzuki:2011hu}, Baryon Acoustic Oscillations (BAO)
\cite{Blake:2011en,Percival:2009xn,Beutler:2011hx,Jarosik:2010iu},
and CMB shift parameter \cite{Komatsu:2010fb} and Hubble parameter data
\cite{Simon:2004tf,Stern:2009ep,Moresco:2012by,Zhang:2012mp,Blake12,Busca12,Chuang2012b}.

Since we have many parameters we fix some them from the stability
conditions. It is clear from Table~\ref{qd} that the fixed points
and their conditions for stability do not depend on $\alpha_3$ and
$\alpha_4$. Thus the cosmology does  not depend on the $\alpha$'s
and consequently we can not get any constraint on $\alpha$'s. But we
can have some conditions on $\alpha$'s from the consideration that
at the fixed point $\Omega_\Lambda>0$. From these conditions we can
fix some values for $\alpha_3$ and $\alpha_4$. One should note that
maintaining the conditions on $\alpha$'s if one change their values
then the cosmology does not change and that is why we can not have
any constraint on the $\alpha$'s \cite{Gannouji:2013rwa}. So we fix
$\alpha_3=1$ and $\alpha_4=0.115$ which respect the conditions on
$\alpha$'s for the point $P_{6+}$. We also fix the present value of
the Hubble parameter $H_0$ from the Planck 2013 result
\cite{Ade:2013zuv}. Since the autonomous system and the fixed points
do not depend on the value of $C$, we can not also derive any
constraint on it. Also the numerical evolution of the autonomous
system does not depend on graviton mass $m$ but we can have an
estimation of the value of $m$ from the evolution of the Hubble
parameter for the best fit values of the other model parameters. So
finally we are left with three model parameters $\beta_0$,
$\omega_0$ and $\Omega_{m0}$.

\begin{figure}[ht]
\begin{center}
\mbox{\epsfig{figure=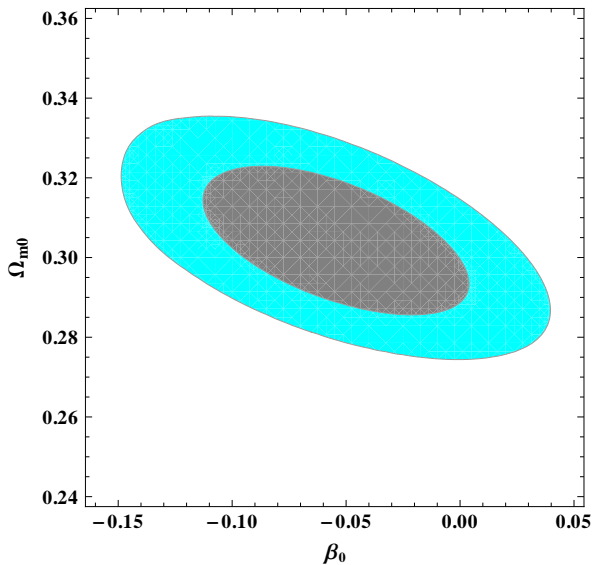,width=7.8cm,angle=0}}
\caption{The 1$\sigma$ and 2$\sigma$ likelihood
contours in the $\beta_0-\Omega_{m0}$ plane are shown here for the parameter choices $\alpha_3=1$, $\alpha_4=0.115$.
Gray and Cyan colors are representing the 1$\sigma$ and 2$\sigma$ contours respectively.}
\label{obs_omega}
\end{center}
\end{figure}

\begin{figure}[ht]
\begin{center}
\mbox{\epsfig{figure=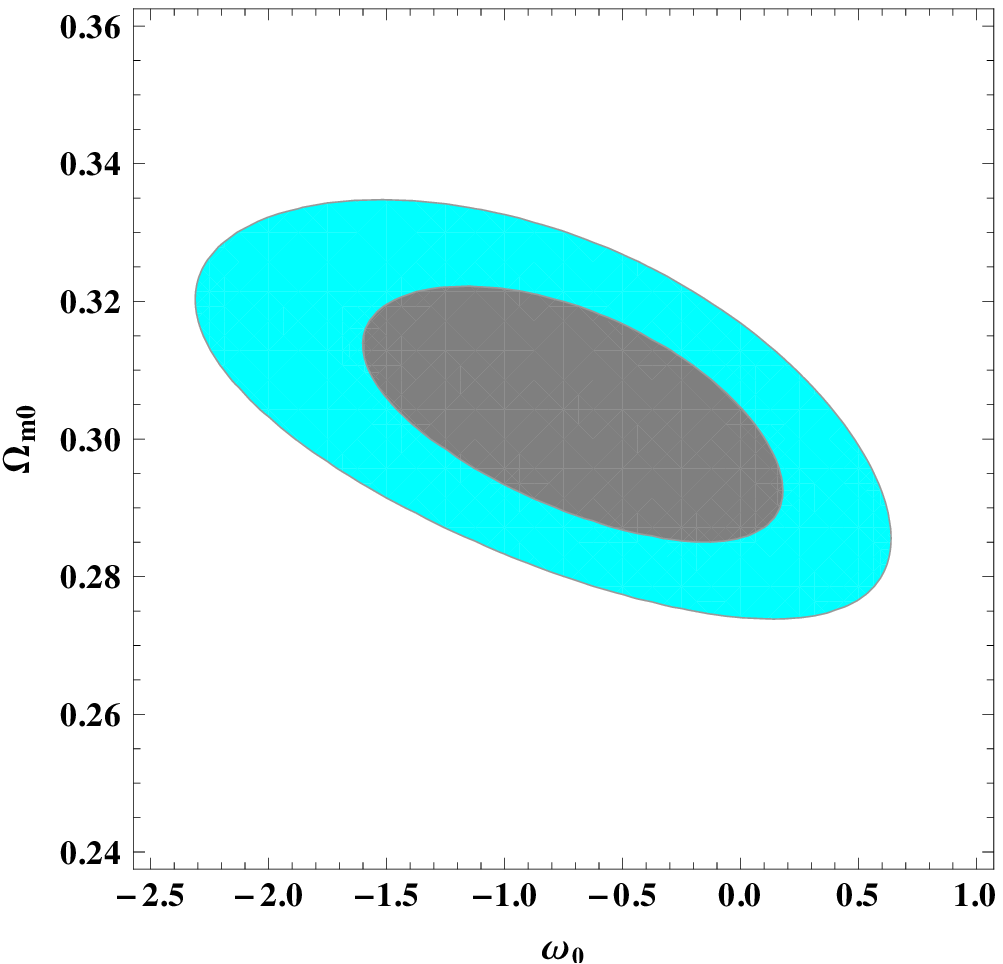,width=7.8cm,angle=0}}
\caption{The 1$\sigma$ and 2$\sigma$ likelihood
contours in the $\omega_0-\Omega_{m0}$ plane are shown here for the parameter choices $\alpha_3=1$, $\alpha_4=0.115$.
Gray and Cyan colors are representing the 1$\sigma$ and 2$\sigma$ contours respectively.}
\label{obs_beta}
\end{center}
\end{figure}

\begin{figure}[ht]
\begin{center}
\mbox{\epsfig{figure=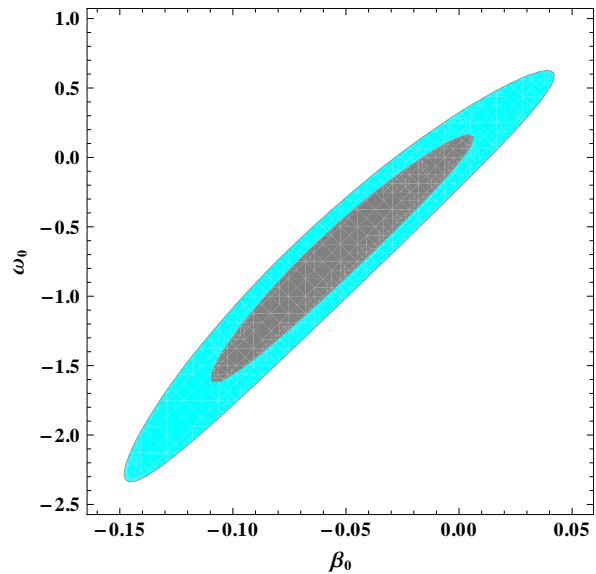,width=7.8cm,angle=0}}
\caption{The 1$\sigma$ and 2$\sigma$ likelihood
contours in the $\beta_0-\omega_0$ plane are shown here for the parameter choices $\alpha_3=1$, $\alpha_4=0.115$.
Gray and Cyan colors are representing the 1$\sigma$ and 2$\sigma$ contours respectively.}
\label{obs_Omega}
\end{center}
\end{figure}

Total $\chi^2$ is defined in Eq.~(\ref{chi_tot}). Along with the
total $\chi^2$ which incorporates  the recent Planck result, we have
also used the estimation on the present value of matter density
parameter ($\Omega_{m0}$) as a prior information where mean value of
$\Omega_{m0}$ lies at 0.315. We have taken the maximum uncertainty
of $\Omega_{m0}$ measurement in Planck results which is 0.018. The
detailed procedure of data analysis is given in Appendix. We have
varied the parameter $\beta_0$ from $-0.2$ to $0.1$, parameter
$\omega_0$ from $-2.5$ to $1$ and $\Omega_{m0}$ from $0.24$ to
$0.36$. Minimizing the total $\chi^2$, we get the best fit values
for the parameters. For this model best fit values of the considered
model parameters are $\beta_0 \approx -0.04$, $\omega_0 \approx -0.5$
and $\Omega_{m0} \approx 0.3$.

Fig.~\ref{obs_omega} shows the 1$\sigma$ and 2$\sigma$ contours on
$\beta_0-\Omega_{m0}$ plane after marginalizing on the parameter
$\omega_0$. Similarly Figs.~\ref{obs_beta} and \ref{obs_Omega}
show the 1$\sigma$ and 2$\sigma$ contours on
$\omega_0-\Omega_{m0}$ and $\beta_0-\omega_0$ planes respectively
after marginalizing on $\beta_0$ and $\Omega_{m0}$ respectively.
Fig.~\ref{obs_omega} shows that very small value of $\beta_0$ is
allowed. Fig.~\ref{obs_beta} shows the consistency with the results
depicted in Table~\ref{qd} which tells us that very small positive
value of $\omega_0$ is allowed to have viable thermal history of the
universe. This figure also tells us that negative values of
$\omega_0$ can also give viable cosmology corresponding to phantom
phase in the evolution of the dark energy component.

Next from Eq.~(\ref{beta0}), we observe that if we put the conformal
coupling constant $\beta=0$ then $\beta_0$ also becomes zero and the
theory reduces to the original quasi-dilaton nonlinear massive
gravity \cite{D'Amico:2012zv}. But observationally we found that
though $\beta_0=0$ is allowed within the $1\sigma$ confidence level
we also have some non zero values of $\beta_0$ allowed within the
$1\sigma$ confidence level; the best fit value of $\beta_0$ is non
zero. Given the present data. However, given the present data, we
can not claim which one of them, minimal/non minimal coupling is
more favored. We find that generally small numerical values of non
minimal coupling in the quasi-dilaton nonlinear massive gravity are
favored.

\section{Conclusions}
\label{conclusions}
In this paper, we  have investigated  the effect of non-minimal
coupling in (extended) massive gravity theories. As our first
example,  we have studied the case of mass varying nonlinear massive
gravity in Brans-Dicke  framework. In this scenario, the conformal
coupling is responsible for making the graviton mass varying. General
theory of relativity is restored  by Vainshtein screening in this
case such that conformal coupling does not affect the local physics.

Our analysis shows that if we do not complement the $\sigma$
Lagrangian by potential or by a non linear  galileon Lagrangian, the
conformal coupling alone can not give rise to de Sitter solution. It
gives rise to an effective pressure of matter such that the matter
dominated era is a late time attractor. And as a result we do not
have the required equation of state at late times.

We next invoke cubic galileon Lagrangian and potential for the
$\sigma$ field. Adding cubic galileon alone leads to a de Sitter
solution as an attractor of the dynamics but simultaneously we also
have matter dominated era as an attractor solution. We have
bi-stability in this case and while following the thermal history of
the universe, we can not reach the de Sitter phase because the
latter is blocked by the matter era. Only by invoking unnatural fine
tunings, we can realize the de Sitter phase which is not desirable.
However, replacing the galileon term by a potential of the $\sigma$
field, it is possible to obtain a stable de Sitter solution. It is
interesting that the potential term turns the matter phase into a
saddle point under certain conditions. Though this scenario can give
us a  right sequence of different cosmological eras, we can not get
rid of the problem of having a constant non zero equation of state
for the matter phase. Evolution of the normalized Hubble parameter
in this model is compared to that of the minimal case and the
$\Lambda$CDM using the available data and it is found that the
evolution of the Hubble parameter in this model does not fit with
data  while the case of minimal coupling  is consistent with
observations. Thus the conformal coupling appreciably changes the
evolution of the Hubble parameter.

Next we  studied the effect of conformal coupling for quasi-dilaton
nonlinear massive gravity. In this case we have considered an extra
coupling $\beta_q$ which makes the model consistent in the limit
$\beta\to 0$. We have performed detailed dynamical analysis for this
model. In this setting, we find a de Sitter solution as a late time
attractor. We have shown that model under consideration gives rise
to viable cosmology. From the dynamical analysis, it is found that
the fixed points and their conditions for stability do not depend
upon the numerical values of the parameters, $\alpha_3$ and
$\alpha_4$ thereby cosmology is not sensitive to  these parameters.
Also we do not find  any dependency of cosmological dynamics on the
graviton mass $m$ and parameter $C$ as these two parameters do not
appear in the dynamical system explicitly.

We have also analyzed the observational constraints on the model
parameters $\beta$, $\omega$ and $\Omega_{mo}$. It is found that
both the parameters $\beta$ and $\omega$ are preferred to have small
numerical values both positive and negative allowed within the
2$\sigma$ confidence level.

Our investigations reveal that not any mass-varying non linear
massive gravity scheme can give rise to a viable cosmology at the
background level. The quasi-dilaton model turns out to be very
generic in the list of such models. And our analysis shows that the
non minimal coupling in the quasi-dilaton nonlinear massive gravity
can give rise to viable cosmology. From the observational data
analysis it is found that small value of the conformal coupling is
favored. It is remarkable that the introduction of non-minimal
coupling in quasi-dilaton model can give rise to a transient phantom
phase allowed by the recent data with certain preference \cite{Ade:2013zuv}. This  
is a generic feature of the scenario that allows to
realize phantom behavior without a phantom field. Future
observations might distinguish the quasi-dilaton model from its
counter part with non minimal coupling considered in this paper.

Last but not least, It will be important to investigate
whether the non-minimal coupling alone (without the extra terms
introduced in \cite{DeFelice:2013tsa} which do not disturb background) can ensure the
stability of the background. If it does not, it will be important to
incorporate the extra terms along with non minimal coupling and
investigate the phantom behavior and stability issues. We defer this
work for our future investigations.


\begin{acknowledgments}
\noindent We thank T. Koivisto for useful comments. M.W.H.
acknowledges the funding from CSIR, govt. of India. The work is
supported in part by the JSPS Grant-in-Aid for Young Scientists (B)
\# 25800136 (K.B.) and that for Scientific Research (S) \# 22224003
and (C) \# 23540296 (S.N.). M.S. thanks Kobayashi-Maskawa Institute
(KMI) for the Origin of Particles and the Universe for hosting him
as a senior JSPS fellow (L-12522), where the work was initiated. He
also thanks the Eurasian International Center for Theoretical
Physics, Eurasian National University, Astana 010008, Kazakhstan,
 where the work was completed. R.M. thanks KMI for hospitality during his visit
 to Nagoya, during KMI-IEEC Joint International Workshop
-- Inflation, Dark Energy, and Modified Gravity in the PLANCK Era. He
acknowledges fruitful discussions with S. Nojiri and K. Bamba on the related
theme.
\end{acknowledgments}

\appendix*

\begin{table*}[!]
\caption{Values of $\frac{d_A(z_\star)}{D_V(Z_{BAO})}$ for different values
of $z_{BAO}$.}
\begin{center}
\resizebox{\textwidth}{!}{%
\begin{tabular}{c||cccccc}
\hline\hline
 $z_{BAO}$  & 0.106~~~  & 0.2~~~ & 0.35~~~ & 0.44~~~ & 0.6~~~ & 0.73\\
\hline
 $\frac{d_A(z_\star)}{D_V(Z_{BAO})}$ &  $30.95 \pm 1.46$~~~ & $17.55 \pm 0.60$~~~
& $10.11 \pm 0.37$ ~~~& $8.44 \pm 0.67$~~~ & $6.69 \pm 0.33$~~~ & $5.45 \pm 0.31$
\\
\hline\hline
\end{tabular}}
\label{baodata}
\end{center}
\end{table*}

\section{Observational Data Analysis}
\label{Observ}

Here we briefly review the sources of observational constraints used in
this manuscript, namely Type Ia Supernovae constraints, BAO, CMB and data of Hubble parameter. The total $\chi^2$ defined as
\begin{align}
\chi_{\rm tot}^2=\chi_{\rm SN}^2+\chi_{\rm Hub}^2+\chi_{\rm BAO}^2+\chi_{\rm CMB}^2 \,,
\label{chi_tot}
\end{align}
where the individual $\chi^2_i$ for every data set is calculated as
follows.   \\

\begin{table}
\caption{Hubble parameter versus redshift data \cite{Farooq:2013hq}.}
\begin{center}
\label{hubble}
\begin{tabular}{cccc}
\hline\hline
~~$z$ & ~~~~$H(z)$ &~~~~ $\sigma_{H}$ & ~~ Reference\\
~~    &~~ (km/s/Mpc) &~~~~ (km/s/Mpc)& \\
\tableline
0.070&~~    69&~~~~~~~  19.6&~~ \cite{Zhang:2012mp}\\
0.100&~~    69&~~~~~~~  12&~~   \cite{Simon:2004tf}\\
0.120&~~    68.6&~~~~~~~    26.2&~~ \cite{Zhang:2012mp}\\
0.170&~~    83&~~~~~~~  8&~~    \cite{Simon:2004tf}\\
0.179&~~    75&~~~~~~~  4&~~    \cite{Moresco:2012by}\\
0.199&~~    75&~~~~~~~  5&~~    \cite{Moresco:2012by}\\
0.200&~~    72.9&~~~~~~~    29.6&~~ \cite{Zhang:2012mp}\\
0.270&~~    77&~~~~~~~  14&~~   \cite{Simon:2004tf}\\
0.280&~~    88.8&~~~~~~~    36.6&~~ \cite{Zhang:2012mp}\\
0.350&~~    76.3&~~~~~~~    5.6&~~  \cite{Chuang2012b}\\
0.352&~~    83&~~~~~~~  14&~~   \cite{Moresco:2012by}\\
0.400&~~    95&~~~~~~~  17&~~   \cite{Simon:2004tf}\\
0.440&~~    82.6&~~~~~~~    7.8&~~  \cite{Blake12}\\
0.480&~~    97&~~~~~~~  62&~~   \cite{Stern:2009ep}\\
0.593&~~    104&~~~~~~~ 13&~~   \cite{Moresco:2012by}\\
0.600&~~    87.9&~~~~~~~    6.1&~~  \cite{Blake12}\\
0.680&~~    92&~~~~~~~  8&~~    \cite{Moresco:2012by}\\
0.730&~~    97.3&~~~~~~~    7.0&~~  \cite{Blake12}\\
0.781&~~    105&~~~~~~~ 12&~~   \cite{Moresco:2012by}\\
0.875&~~    125&~~~~~~~ 17&~~   \cite{Moresco:2012by}\\
0.880&~~    90&~~~~~~~  40&~~   \cite{Stern:2009ep}\\
0.900&~~    117&~~~~~~~ 23&~~   \cite{Simon:2004tf}\\
1.037&~~    154&~~~~~~~ 20&~~   \cite{Moresco:2012by}\\
1.300&~~    168&~~~~~~~ 17&~~   \cite{Simon:2004tf}\\
1.430&~~    177&~~~~~~~ 18&~~   \cite{Simon:2004tf}\\
1.530&~~    140&~~~~~~~ 14&~~   \cite{Simon:2004tf}\\
1.750&~~    202&~~~~~~~ 40&~~   \cite{Simon:2004tf}\\
2.300&~~    224&~~~~~~~ 8&~~    \cite{Busca12}\\

\hline\hline
\end{tabular}
\end{center}
\end{table}

{\it{a. $\chi^2$ for Type Ia Supernovae}}\\

In order to incorporate Type Ia constraints  we use the Union2.1 data
compilation \cite{Suzuki:2011hu} of 580 data points. The relevant
observable is the distance modulus $\mu$ which is defined as  $\mu=m - M=5
\log D_L+\mu_0$,
where $m$ and $M$ are the apparent and absolute
magnitudes of the Supernovae, $D_L(z)$ is the luminosity distance
$D_L(z)=(1+z)
\int_0^z\frac{H_0dz'}{H(z')}$ and
$\mu_0=5 \log\left(\frac{H_0^{-1}}{\rm Mpc}\right)+2
5$ is a nuisance parameter that should be marginalized. The
corresponding $\chi^2$ writes as
\begin{align}
\chi_{\rm SN}^2(\mu_0,\theta)=\sum_{i=1}^{580}
\frac{\left[\mu_{th}(z_i,\mu_0,\theta)-\mu_{obs}(z_i)\right]^2}{
\sigma_\mu(z_i)^2}\,,
\end{align}
where $\mu_{obs}$ denotes the observed distance modulus while $\mu_{th}$
the theoretical one, and $\sigma_{\mu}$ is the uncertainty in the distance
modulus. Additionally, $\theta$ denotes any parameter of the specific
model at hand. Finally, marginalizing $\mu_0$ following
\cite{Lazkoz:2005sp} we obtain
\begin{align}
\chi_{\rm SN}^2(\theta)=A(\theta)-\frac{B(\theta)^2}{C(\theta)}\,,
\end{align}
with
\begin{align}
&A(\theta) =\sum_{i=1}^{580}
\frac{\left[\mu_{th}(z_i,\mu_0=0,\theta)-\mu_{obs}(z_i)\right]^2}{
\sigma_\mu(z_i)^2}\,, \\
&B(\theta) =\sum_{i=1}^{580}
\frac{\mu_{th}(z_i,\mu_0=0,\theta)-\mu_{obs}(z_i)}{\sigma_\mu(z_i)^2}\,, \\
&C(\theta) =\sum_{i=1}^{580} \frac{1}{\sigma_\mu(z_i)^2}\,.
\end{align}

{\it{d. $\chi^2$ for Hubble data}}\\

Along with the present value of the Hubble parameter we have 29 data points of Hubble parameter given in Table~\ref{hubble}.
To use the data of Hubble parameter we have used the normalized Hubble parameter defined as,
$h=H/H_0$. We fix the value of $H_0$ from \cite{Ade:2013zuv}. Error in normalized Hubble parameter is given by,
\begin{equation}
 \sigma_h=\(\frac{\sigma_H}{H}+\frac{\sigma_{H_0}}{H_0}\) h \,,
 \label{errorh}
\end{equation}
where $\sigma_H$ is the error in Hubble parameter and $\sigma_{H_0}$ is the error in present value of the Hubble parameter.

The $\chi^2$ for the normalized Hubble parameter data is given by,
\begin{align}
\chi_{\rm Hub}^2(\mu_0,\theta)=\sum_{i=1}^{29}
\frac{\left[h_{\rm th}(z_i,\theta)-h_{\rm obs}(z_i)\right]^2}{
\sigma_H(z_i)^2}\,,
\end{align}
where $h_{\rm th}$ and $h_{\rm obs}$ are the theoretical and observed values of the normalized Hubble parameter.

{\it{b. $\chi^2$ for Baryon Acoustic Oscillation (BAO)}}\\

We use BAO data from
\cite{Blake:2011en,Percival:2009xn,Beutler:2011hx,Jarosik:2010iu},
that is of $\frac{d_A(z_\star)}{D_V(Z_{BAO})}$,
where $z_\star$ is the decoupling time given by $z_\star \approx 1091$, the
co-moving angular-diameter distance $d_A(z)=\int_0^z \frac{dz'}{H(z')}$
and
$D_V(z)=\left(d_A(z)^2\frac{z}{H(z)}\right)^{\frac{1}{3}}$
is the dilation scale \cite{Eisenstein:2005su}. th required data
are depicted in Table~\ref{baodata}.

In order to calculate $\chi_\mathrm{BAO}^2$ for BAO data
we follow the procedure
described in Ref.~\cite{Giostri:2012ek}, where it is defined as,
\begin{equation}
 \chi_{\rm BAO}^2=X_{\rm BAO}^T C_{\rm BAO}^{-1} X_{\rm BAO}\,,
\end{equation}
with
\begin{equation}
X_{\rm BAO}=\left( \begin{array}{c}
        \frac{d_A(z_\star)}{D_V(0.106)} - 30.95 \\
        \frac{d_A(z_\star)}{D_V(0.2)} - 17.55 \\
        \frac{d_A(z_\star)}{D_V(0.35)} - 10.11 \\
        \frac{d_A(z_\star)}{D_V(0.44)} - 8.44 \\
        \frac{d_A(z_\star)}{D_V(0.6)} - 6.69 \\
        \frac{d_A(z_\star)}{D_V(0.73)} - 5.45
        \end{array} \right)\,,
\end{equation}
and the inverse covariance matrix reads as
\begin{widetext}
\begin{align}
C^{-1}=\left(
\begin{array}{cccccc}
 0.48435 & -0.101383 & -0.164945 & -0.0305703 & -0.097874 & -0.106738 \\
 -0.101383 & 3.2882 & -2.45497 & -0.0787898 & -0.252254 & -0.2751 \\
 -0.164945 & -2.45499 & 9.55916 & -0.128187 & -0.410404 & -0.447574 \\
 -0.0305703 & -0.0787898 & -0.128187 & 2.78728 & -2.75632 & 1.16437 \\
 -0.097874 & -0.252254 & -0.410404 & -2.75632 & 14.9245 & -7.32441 \\
 -0.106738 & -0.2751 & -0.447574 & 1.16437 & -7.32441 & 14.5022
\end{array}
\right)\,.
\end{align}
\end{widetext}

{\it{c. $\chi^2$ for CMB shift parameter}}\\

We use the CMB shift parameter
\begin{align}
R=H_0 \sqrt{\Omega_{m0}}
\int_0^{1089}\frac{dz'}{H(z')}\,,
\end{align}
 following \cite{Komatsu:2010fb}.
The corresponding $\chi_{\rm CMB}^2$ is defined as,
\begin{align}
 \chi_{\rm CMB}^2(\theta)=\frac{(R(\theta)-R_0)^2}{\sigma^2}\,,
\end{align}
with $R_0=1.725 \pm 0.018$ and $R(\theta)$ \cite{Komatsu:2010fb}.

\end{document}